\documentclass[%
 reprint,
 amsmath,amssymb,
 aps,
prb,
floatfix,
]{revtex4-2}

\usepackage{dcolumn}
\usepackage{bm, amsmath, amsfonts, amssymb, mathtools, braket}
\usepackage{multirow}
\usepackage[utf8]{inputenc}
\usepackage{graphicx} 
\usepackage{float, xcolor}
\usepackage{physics}
\usepackage{comment}
\usepackage{mathrsfs}

\usepackage[
pagebackref=false,
colorlinks=true,
linkcolor=black,
urlcolor=blue,
filecolor=black,
citecolor=blue,
pdfstartview=FitV,
pdftitle={},
pdfauthor={},
pdfsubject={},
pdfkeywords={},
bookmarksopen=true
]{hyperref}

\newcommand{\1}{\mbox{1}\hspace{-0.25em}\mbox{l}}
\newcommand{\ii}{\mathrm{i}}

\begin{document}


\title{Liouvillian skin effects in two-dimensional electron systems at finite temperatures}

\author{Yuta Shigedomi}
\email{shigedomi.yuta.37p@st.kyoto-u.ac.jp}
\author{Tsuneya Yoshida}%
\email{yoshida.tsuneya.2z@kyoto-u.ac.jp}
\affiliation{%
 Department of Physics, Kyoto University, Kyoto 606-8502, Japan
}%
\date{\today}

\begin{abstract}
Liouvillian skin effects, manifested as the localization of Liouvillian eigenstates around the boundary, are distinctive features of non-Hermitian systems and are particularly notable for their impact on system dynamics. Despite their significance, Liouvillian skin effects have not been sufficiently explored in electron systems. 
In this work, we demonstrate that a two-dimensional electron system on a substrate exhibits $\mathbb{Z}$ and $\mathbb{Z}_2$ Liouvillian skin effects due to the interplay among energy dissipations, spin-orbit coupling, and a transverse magnetic field.
In addition, our analysis of the temperature dependence reveals that these Liouvillian skin effects become pronounced below the energy scale of band splitting induced by the spin-orbit coupling and the magnetic field.
While our $\mathbb{Z}$ Liouvillian skin effect leads to charge accumulation under quench dynamics, its relaxation time is independent of the system size, in contrast to that of previously reported Liouvillian skin effects. This difference is attributed to the scale-free behavior of the localization length, which is analogous to non-Hermitian critical skin effects.
\end{abstract}

\maketitle

\section{INTRODUCTION}

In recent years, non-Hermitian topology has attracted growing interest, as it induces a variety of exotic phenomena that have no Hermitian counterparts~\cite{Kozii_PRB2024,Gong_PRX2018,Shen_PRL2018,Yao_PRL2018,Yoshida_PRB2018,Yoshida_PRB2019,Budich_PRB2019,Okugawa_PRB2019,Kawabata_PRX2019,Yokomizo_PRL2019,Ashida_AiP2020,Bergholtz_RMP2021}.
A prime example is the non-Hermitian skin effect; due to the point gap topology unique to non-Hermitian systems, both eigenvalues and eigenstates exhibit extreme sensitivity to boundary conditions~\cite{Kunst_PRL2018,Yao_Wang_PRL2018,Lee_PRB2019,Longhi_PRR2019,Okuma_PRL2020,Borgnia_PRL2020,Zhang_PRL2020,Kawabata_PRB2020,Okugawa_PRB2020,Li_NatComm2020,Yokomizo_PRB2021,Okuma_PRB2021,Franca_PRL2022,Zhang_AiPX2022,Zhou_PRA2022,Lin_FoP2023,Kawabata_PRX2023,Qin_PRB2023,Hwang_PRB2023,Okuma_AnnRevCondMatPhys2023,Yoshida_PRL2024,Shimomura_PRL2024,Mandal_JoAP2024,Ji_arXiv2025,Nakagawa_PRX2025}. 
Specifically, in one-dimensional systems with nonreciprocal hopping, most eigenstates are localized around the boundary under open boundary conditions. In addition, symmetry protection of point gap topology enriches the skin effect; for example, time-reversal symmetry protects the $\mathbb{Z}_2$ skin effect~\cite{Okuma_PRL2020,Wan_PRL2023,Kaneshiro_PRB2023,Ishikawa_PRB2024,Tanaka_Okugawa_PRB2024}. 
The non-Hermitian skin effects have been experimentally observed
for a wide range of platforms~\cite{Ochkan_NatPhys2024} from metamaterials~\cite{Ghatak_PNAS2020,Weidemann_Science2020,Helbig_NatPhys2020,Hofmann_PRR2020,Zhang_Natcomm2021,Li_NatComm2024} to synthetic quantum systems such as cold atoms~\cite{Liang_PRL2022,Zhao_Nat2025}.

Notably, non-Hermitian skin effect has been extended to the Liouvillian of the Gorini–Kossakowski–Sudarshan–Lindblad (GKSL) equation describing open quantum systems~\cite{Lindblad_CiM1976,Gorini_JoMP1976,Breuer_OUP2002}.
The Liouvillian skin effects significantly affect dynamical properties~\cite{Song_PRL2019,Liu_PRR2020,Mori_PRL2020,Haga_PRL2021,Yang_PRR2022,Hamanaka_PRB2023,Li_PRR2023,Wang_PRB2023,Begg_PRL2024,Liu_PRL2024,Ekman_2024PRR,Feng_PRB2024,Feng_PRB2024,Hamanaka_PRB2025,Cai_PRB2025,Kuo_PRR2025,Li_arXiv2025,Pena_PRB2025}. 
The localization of Liouvillian eigenstates around the boundary induces spatially inhomogeneous dynamics of the particle number and leads to system-size dependent relaxation times, which can diverge in the thermodynamic limit.

Despite the importance of Liouvillian skin effects governing dynamical properties, they remain largely unexplored in electron systems of solids. 
In particular, temperature effects -- ubiquitous in solids -- have not been thoroughly investigated; if temperature effects suppress the skin effects, how does this suppression persist under the dissipations that would exist in high-temperature regions?

In this work, we demonstrate that a two-dimensional electron system on a substrate exhibits $\mathbb{Z}$ and $\mathbb{Z}_2$ Liouvillian skin effects due to the interplay among energy dissipations, Rashba spin-orbit coupling (SOC), and a transverse magnetic field. 
In our electron system, these Liouvillian skin effects become pronounced 
below the energy scale of band splitting induced by the SOC and the magnetic field; in the high temperature regions, dissipations become homogeneous in momentum space, suppressing the skin effects.
While our $\mathbb{Z}$ Liouvillian skin effect leads to 
charge accumulation under quench dynamics, its relaxation time is independent of the system size, in contrast to that of previously reported Liouvillian skin effects. This difference is attributed to the scale-free behavior of the localization length for our system, which is analogous to non-Hermitian critical skin effects.
\\

The rest of this paper is organized as follows. Section~\ref{Section:Model} introduces our two-dimensional electron system. In Sec.~\ref{Section:Methods}, we outline our approach of analysis and briefly review relevant topological invariants. Section~\ref{Section:Results} provides eigenstates and eigenvalues of our Liouvillian exhibiting the extreme sensitivity, as well as analyses of temperature effects and dynamical properties. Section~\ref{Section:Conclusion} provides a brief summary.

The Appendixes provide the details of the analytical methods and a discussion of the non-Hermitian critical skin effects of the Liouvillian. They further provide a classification of the skin effects for general antisymmetric SOC and Zeeman fields, a remark on the instability of the $\mathbb{Z}_2$ skin modes under symmetry-breaking perturbations, and an analysis of the Liouvillian skin effect under fully open boundary conditions.

\section{Model\label{Section:Model}}
\begin{figure}[htbp]
\includegraphics[width=0.48\textwidth]{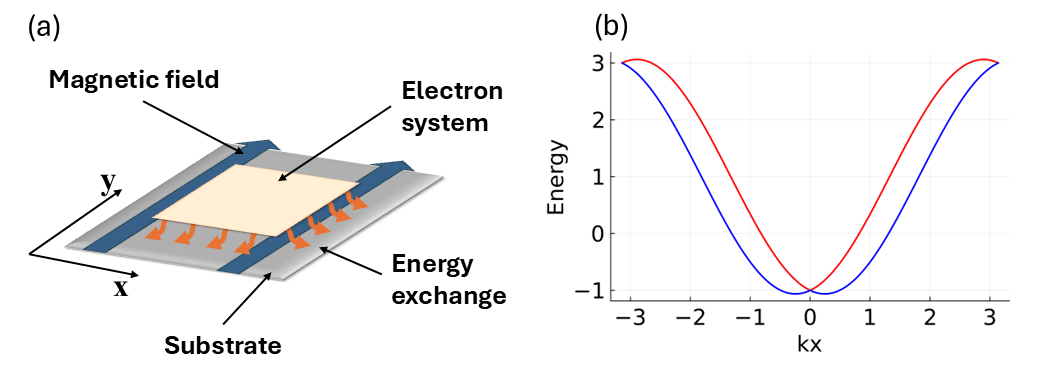}
\caption{\label{Figure:setup}
(a) Schematic illustration of the system: a two-dimensional electron system on a substrate under an in-plane magnetic field, with the energy exchange (dissipation) between the system and the substrate.
(b) 
Energy eigenvalues of the Hamiltonian [Eq.~\eqref{eq: Hamiltonian}] at $k_y=\pi$.
The red and blue curves represent $\varepsilon_{\vb*{k}1}$ and $\varepsilon_{\vb*{k}2}$, respectively.
The band splitting is induced by the Rashba SOC.
The data are obtained for $t_{\mathrm{h}} = 1$, $\mu = 1$, $\alpha = 0.5$, $H_x = H_y = 0$, and $k_y = \pi$.
}
\end{figure}

We consider a two-dimensional electron system on a square lattice placed on a substrate [Fig.~\ref{Figure:setup}(a)]~\cite{experimentalNote,Bollinger_Nature2011,Leng_PRL2011}. We take into account energy exchange (or energy dissipation) between the system and the substrate. Markovian dynamics of the system is described by the GKSL equation which is given by
\begin{eqnarray}
\ii\dv{\rho}{t} 
&=& [H,\rho] 
   + \ii\sum_{lm}\qty[
      L_{lm}\rho L_{lm}^\dagger 
      - \frac{1}{2}\Big\{L_{lm}^\dagger L_{lm},\rho\Big\}
   ]\\
&\equiv& \mathscr{L}(\rho).
\nonumber
\end{eqnarray}

Here, $\mathscr{L}$ denotes the Liouvillian which is a superoperator acting on the density matrix of the system $\rho$. The Hamiltonian of the system is denoted by the Hermitian operator $H$. The jump operators $ L_{lm}$ describe the coupling to the substrate. The indices $l$ and $m$ label relaxation processes.
In the following, we provide the details of Hamiltonian $H$ and jump operators $L_{lm}$.

\subsection{Hamiltonian}
The Hamiltonian under a magnetic field reads
\begin{equation}
\label{eq: Hamiltonian}
    H = \sum_{\vb*{k}\sigma}\xi_{\vb*{k}}c_{\vb*{k}\sigma}^\dagger c_{\vb*{k}\sigma}
    +
    \sum_{\vb*{k}\sigma\sigma'}(\alpha\vb*{g}_{\vb*{k}}-\mu_{\mathrm{B}}\vb*{H})\vdot \vb*{\sigma}_{\sigma\sigma'}c_{\vb*{k}\sigma}^\dagger c_{\vb*{k}\sigma'},
\end{equation}
where $c_{\vb*{k}\sigma}^\dagger$ ($c_{\vb*{k}\sigma}$) is the creation (annihilation) operator 
of an electron with momentum $\vb*{k}=(k_x,k_y)$ and spin 
$\sigma=\uparrow,\downarrow$, and 
$\vb*{\sigma}=(\sigma_x,\sigma_y,\sigma_z)$ is the vector of Pauli matrices acting on the spin space.

The first term describes the kinetic energy of the electrons with $\xi_{\vb*{k}} = -2t_{\mathrm{h}}(\cos{k_x} + \cos{k_y}) - \mu$, which does not act on the spin.
Here, $t_{\mathrm{h}}$ denotes the nearest-neighbor hopping, and $\mu$ is the chemical potential.

The second term describes Rashba SOC arising from local reflection symmetry breaking and the Zeeman term. The Rashba SOC is represented by the momentum-dependent vector 
$\vb*{g}_{\vb*{k}} = (-\sin{k_y},\sin{k_x},0)$
 with coupling strength denoted by $\alpha$. The Zeeman term involves the Bohr magneton $\mu_{\mathrm{B}}$ and
the in-plane magnetic field
$\vb*{H} = (H_x,H_y,0)$
.
 We can obtain the energy dispersion of the Hamiltonian via the diagonalization in each $\vb*{k}$ subspace
\begin{equation}
    H =
    \sum_{\vb*{k}} \vb{C}_{\vb*{k}}^\dagger
    \begin{pmatrix}
        \xi_{\vb*{k}}&\eta_{\vb*{k}}\\
        \eta_{\vb*{k}}^*&\xi_{\vb*{k}}
    \end{pmatrix}
    \vb{C}_{\vb*{k}}
    =
    \sum_{\vb*{k}} 
    \vb{C}{'}_{\vb*{k}}^{\dagger}
    \begin{pmatrix}
        \varepsilon_{\vb*{k}1}&0\\
        0&\varepsilon_{\vb*{k}2}
    \end{pmatrix}
    \vb{C}{'}_{\vb*{k}},
\end{equation}
with
$\eta_{\vb*{k}}= -\alpha\sin{k_y}-\mu_BH_x -\ii(\alpha\sin{k_x}-\mu_BH_y)$
, $\vb{C}_{\vb*{k}}=\begin{pmatrix}c_{\vb*{k}\uparrow }&c_{\vb*{k}\downarrow }\end{pmatrix}^T,
$
and
$
\vb{C}{'}_{\vb*{k}}=\begin{pmatrix}\alpha_{\vb*{k}1}&\alpha_{\vb*{k}2 }\end{pmatrix}^T
$.

The energy eigenvalues are given by
\begin{equation}
\begin{split}
    \label{Equation:band}    
    &\varepsilon_{\vb*{k}1}=\xi_{\vb*{k}}+\abs{\eta_{\vb*{k}}},\\ 
    &\varepsilon_{\vb*{k}2}=\xi_{\vb*{k}}-\abs{\eta_{\vb*{k}}},\\
    &\abs{\eta_{\vb*{k}}}=\sqrt{ (\alpha\sin{k_y} + \mu_BH_x)^2 + (\alpha\sin{k_x} - \mu_BH_y)^2 },
\end{split}
\end{equation}
and
plotted in Fig.~\ref{Figure:setup}(b).

\subsection{Jump operators}

We consider that the system is coupled to the substrate with energy exchange (or energy dissipation), which conserves the number of electrons. Specifically, we assume the following jump operators~\cite{relaxNote, Ikeda_SciPost2021,Castro_Scipost2023,Tanaka_PRB2024}:

\begin{equation}
L^{\vb*{k}}_{lm} = \sqrt{\Gamma^{\vb*{k}}_{lm}}\big|\varepsilon_{\vb*{k}l}\big\rangle\big\langle\varepsilon_{\vb*{k}m}\big|
=
\sqrt{\Gamma^{\vb*{k}}_{lm}}\alpha_{\vb*{k}l}^\dagger\alpha_{\vb*{k}m},
\end{equation}
where $l$ and $m$ 
denote the band indices. The transition probability $\Gamma^{\vb*{k}}_{lm}$ is defined by
\begin{equation}
    \Gamma^{\vb*{k}}_{lm} = \frac{\gamma_{lm} e^{-\beta\varepsilon_{\vb*{k}l}}}{e^{-\beta\varepsilon_{\vb*{k}l}} + e^{-\beta\varepsilon_{\vb*{k}m}}},
    \quad \gamma_{lm} = 
    \begin{cases}
        \gamma & \text{if $l \neq m$} \\
        \gamma_{\mathrm{d}} & \text{if $l = m$}
    \end{cases},
\end{equation}
with $\beta = 1/T$ denoting the inverse temperature. 
Here, the Boltzmann constant is set to unity throughout this paper. The processes with $l \neq m$ involve energy exchange and are characterized by the rate $\gamma \geq 0$, while those with $l = m$ correspond to energy-conserving processes known as dephasing, with the associated rate denoted by $\gamma_{\mathrm{d}} \geq 0$.

These jump operators satisfy the detailed balance condition~\cite{Breuer_OUP2002}
\begin{equation}
    \frac{\Gamma^{\vb*{k}}_{lm}}{\Gamma^{\vb*{k}}_{ml}} = \frac{e^{-\beta\varepsilon_{\vb*{k}l}}}{e^{-\beta\varepsilon_{\vb*{k}m}}},
\end{equation}
and describe relaxation toward the Gibbs equilibrium state of the Hamiltonian $H$. 

A similar form of jump operators has been employed in previous works to take into account the energy dissipation of Floquet systems~\cite{Ikeda_SA2020, Ikeda_SciPost2021,Castro_Scipost2023,Tanaka_PRB2024}.

\section{Methods\label{Section:Methods}}

We analyze the topology of our Liouvillian in the following steps.
First, we employ a vectorized representation of the density matrix, which maps the superoperator $\mathscr{L}$ to a matrix.
Then, we apply a mean-field approximation to the Liouvillian, which contains quartic terms.
With the obtained matrix, we compute $\mathbb{Z}$ and $\mathbb{Z}_2$ invariants.

\subsection{Vectorization of the density matrix}
We briefly review a vectorized representation of the density matrix. A more detailed discussion is presented in Appendix~\ref{Appendix:properties}.

We begin by mapping the density matrix $\rho$, which acts on the Hilbert space $\mathcal{H}$, to a vector in the doubled Hilbert space $\mathcal{H} \otimes \mathcal{H}$~\cite{
Jamiolkowski_RoMP1972,Choi_LAA1975}
\begin{equation}
    \rho = \sum_{i, j} \rho_{ij} \ket{\phi _{i}}\bra{\phi _{j}} \rightarrow \lvert\rho\rangle\rangle = \sum_{i, j} \rho_{ij} \lvert \phi _{i}\rangle\rangle\otimes \lvert \phi _{j}\rangle\rangle.
    \label{Equation:vectorization}
\end{equation}
We note that the first (second) space of the doubled Hilbert space $\mathcal{H}\otimes\mathcal{H}$ is referred to as the ket (bra) space.

Once the density matrix is vectorized, the Liouvillian $\mathscr{L}$ becomes a linear operator $\mathcal{L}$ acting on a doubled Hilbert space. Its explicit form is 
\begin{equation}
    \begin{split}
        &\mathcal{L} = (H\otimes \1 - \1\otimes H^{*})+ \ii\sum_{lm}(L _{lm}\otimes L _{lm} ^{*} )\\
    &- \frac{\ii}{2}\sum_{lm}\bigg[ \1\otimes \qty(L _{lm}^{\dagger}L _{lm}) ^{*} + L _{lm}^{\dagger}L _{lm}\otimes \1 \bigg].
    \label{Equation:vecGKSL}
    \end{split}       
\end{equation}
The $n$-th eigenvalues $\Lambda_n$ and corresponding right (left) eigenvectors $|\rho_{n}^{\mathrm{R}}\rangle\rangle$ ($\langle\langle\rho_{n}^{\mathrm{L}}|$) are obtained by solving 
\begin{equation}
    \mathcal{L}|\rho_{n}^{\mathrm{R}}\rangle\rangle = \Lambda_{n}|\rho_{n}^{\mathrm{R}}\rangle\rangle,\quad 
    \langle\langle\rho_{n}^{\mathrm{L}}|\mathcal{L} = \langle\langle\rho_{n}^{\mathrm{L}}|\Lambda_{n}
\label{Equation:valvec}
\end{equation}
for $n = 1,\dots,\mathrm{dim} (\mathcal{L})$.

Next, we redefine the fermionic creation and annihilation operators as 
\begin{eqnarray}
    &c_{i\sigma \mathrm{K}}\equiv c_{i\sigma}\otimes \1,\quad c_{i\sigma \mathrm{B}}\equiv \eta (\1\otimes c_{i\sigma}^{*}),
    \label{Equation:KB}\\
    &\eta \equiv  (-1)^{\sum_{i\sigma}n_{i\sigma}}\otimes(-1)^{\sum_{i\sigma}n_{i\sigma}^*},
    \label{Equation:eta}
\end{eqnarray}
following Ref.~\cite{Wang_arXiv2024}. 
Here, $c_{i\sigma \mathrm{K}}$ ($c_{i\sigma \mathrm{B}}$) act on the ket (bra) space.
This definition ensures anticommutation relations between ket and bra operators.
By performing a Fourier transformation, these operators can be expressed in momentum space~\cite{momentumNote}.

Finally, we rewrite the expectation values in the vectorized notation:
\begin{equation}
    \tr(A\rho B) = \langle\langle\1|A\otimes B^{T}|\rho\rangle\rangle.
    \label{Equation:expectationval}
\end{equation}

\subsection{Mean-field approximation} 
We apply a mean-field approximation to our Liouvillian, which includes quartic terms.
Focusing on the near-equilibrium dynamics, we approximate the Liouvillian by using the equilibrium density matrix given by the Gibbs distribution. The relevant expectation values are then evaluated as
\begin{eqnarray}
    &\langle\langle\1|\alpha_{\vb*{k}l \mathrm{K}}^{\dagger}\alpha_{\vb*{k}'m \mathrm{K}}|\rho_{\mathrm{Gibbs}}\rangle\rangle = \delta_{\vb*{k}\vb*{k}'}\delta_{lm}f(\varepsilon_{\vb*{k}l}),\\
    &\langle\langle\1|\alpha_{\vb*{k}l \mathrm{B}}^{\dagger}\alpha_{\vb*{k}'m \mathrm{B}}|\rho_{\mathrm{Gibbs}}\rangle\rangle = \delta_{\vb*{k}\vb*{k}'}\delta_{lm}f(\varepsilon_{\vb*{k}l}),\\
    &\langle\langle\1|\alpha_{\vb*{k}l \mathrm{K}}\alpha_{\vb*{k}'m \mathrm{B}}|\rho_{\mathrm{Gibbs}}\rangle\rangle = -\delta_{\vb*{k}\vb*{k}'}\delta_{lm}f(\varepsilon_{\vb*{k}l}),\label{Equation:scgap1}\\
    &\langle\langle\1|\alpha_{\vb*{k}l \mathrm{K}}^{\dagger}\alpha_{\vb*{k}'m \mathrm{B}}^{\dagger}|\rho_{\mathrm{Gibbs}}\rangle\rangle= \delta_{\vb*{k}\vb*{k}'}\delta_{lm}f(-\varepsilon_{\vb*{k}l}),\label{Equation:scgap2}
\end{eqnarray}
where $f$ denotes the Fermi distribution function. The contribution analogous to the superconducting gap, as expressed in Eqs.~(\ref{Equation:scgap1})~and~(\ref{Equation:scgap2}), naturally remains~\cite{Yamamoto_PRL2021}. By introducing the Nambu basis, which is often used in the analysis of superconductors~\cite{Nambunote},
\begin{equation}
    \Psi_{\vb*{k}} = \begin{pmatrix} \alpha_{\vb*{k}1 \mathrm{K}} & \alpha_{\vb*{k}1 \mathrm{B}}^{\dagger} & 
     \alpha_{\vb*{k}2 \mathrm{K}} & \alpha_{\vb*{k}2 \mathrm{B}}^{\dagger} \end{pmatrix}^T,
\end{equation}
the approximated quadratic Liouvillian $\bar{\mathcal{L}}$ is represented in matrix form
\begin{eqnarray}
    &\bar{\mathcal{L}} = \sum_{\vb*{k}}\Psi_{\vb*{k}}^\dagger \bar{\mathcal{L}}(\vb*{k})\Psi_{\vb*{k}} + \mathrm{const.},\\
    &\bar{\mathcal{L}}(\vb*{k}) = \begin{pmatrix} \bar{\mathcal{L}}_{1}(\vb*{k}) & 0 \\
    0 & \bar{\mathcal{L}}_{2}(\vb*{k}) \end{pmatrix},
    \label{Equation:Liouvillian_op}
\end{eqnarray}
\begin{widetext}
\begin{eqnarray}
    &\bar{\mathcal{L}}_{1}(\vb*{k}) = 
     \begin{pmatrix} 
     \varepsilon_{\vb*{k}1} - \dfrac{\ii}{2}\Big\{\Gamma _{21}^{\vb*{k}} -\gamma f(\varepsilon_{\vb*{k}2})+\dfrac{\gamma_{\mathrm{d}}}{2}\Big\} +\ii\dfrac{\gamma_{\mathrm{d}}}{2} f(\varepsilon_{\vb*{k}1}) & \ii\Gamma_{12}^{\vb*{k}}f(\varepsilon_{\vb*{k}2})+\ii\dfrac{\gamma_{\mathrm{d}}}{2} f(\varepsilon_{\vb*{k}1})\\
     \ii\Gamma_{21}^{\vb*{k}}f(-\varepsilon_{\vb*{k}2}) +\ii\dfrac{\gamma_{\mathrm{d}}}{2} f(-\varepsilon_{\vb*{k}1})& \varepsilon_{\vb*{k}1} + \dfrac{\ii}{2}\Big\{\Gamma _{21}^{\vb*{k}} - \gamma f(\varepsilon_{\vb*{k}2})+\dfrac{\gamma_{\mathrm{d}}}{2}\Big\}-\ii\dfrac{\gamma_{\mathrm{d}}}{2} f(\varepsilon_{\vb*{k}1}) 
     \end{pmatrix}
     ,
     \label{Equation:Liouvillian1}
     \\
     &\bar{\mathcal{L}}_{2}(\vb*{k}) = 
     \begin{pmatrix} \varepsilon_{\vb*{k}2} - \dfrac{\ii}{2}\Big\{\Gamma _{12}^{\vb*{k}} -\gamma f(\varepsilon_{\vb*{k}1})+\dfrac{\gamma_{\mathrm{d}}}{2}\Big\} +\ii\dfrac{\gamma_{\mathrm{d}}}{2} f(\varepsilon_{\vb*{k}2})& \ii\Gamma_{21}^{\vb*{k}}f(\varepsilon_{\vb*{k}1})+\ii\dfrac{\gamma_{\mathrm{d}}}{2} f(\varepsilon_{\vb*{k}2})\\ 
     \ii\Gamma_{12}^{\vb*{k}}f(-\varepsilon_{\vb*{k}1}) +\ii\dfrac{\gamma_{\mathrm{d}}}{2} f(-\varepsilon_{\vb*{k}2})& \varepsilon_{\vb*{k}2} + \dfrac{\ii}{2}\Big\{\Gamma _{12}^{\vb*{k}} - \gamma f(\varepsilon_{\vb*{k}1})+\dfrac{\gamma_{\mathrm{d}}}{2}\Big\} -\ii\dfrac{\gamma_{\mathrm{d}}}{2} f(\varepsilon_{\vb*{k}2})
     \end{pmatrix}.
     \label{Equation:Liouvillian2}
\end{eqnarray}
\end{widetext}
This Liouvillian is analogous to the Bogoliubov–de Gennes Hamiltonian.
Thus, physical observables can be calculated using ``Bogoliubov quasiparticles" (see Appendix~\ref{Appendix:bogoliubov}).

The eigenvalues of the Liouvillian can be obtained analytically as follows:
\begin{equation}
    \begin{split}
        &\Lambda_{\vb*{k}1} = \varepsilon_{\vb*{k}1}\pm \dfrac{\ii}{2}\qty(\dfrac{\gamma}{1 + e^{\beta(\varepsilon_{\vb*{k}2} - \varepsilon_{\vb*{k}1})}}\frac{1 + e^{-\beta\varepsilon_{\vb*{k}1}}}{1 + e^{-\beta\varepsilon_{\vb*{k}2}}} + \dfrac{\gamma_{\mathrm{d}}}{2} ),\\
        &\Lambda_{\vb*{k}2} = \varepsilon_{\vb*{k}2}\pm \dfrac{\ii}{2}\qty(\dfrac{\gamma}{1 + e^{\beta(\varepsilon_{\vb*{k}2} - \varepsilon_{\vb*{k}1})}}\frac{1 + e^{\beta\varepsilon_{\vb*{k}2}}}{1 + e^{\beta\varepsilon_{\vb*{k}1}}} + \dfrac{\gamma_{\mathrm{d}}}{2} ).
        \label{Equation:analytic}
    \end{split}
\end{equation}

The above mean-field approximation satisfies the following properties:
(i) the complete positivity and trace preservation (CPTP) of the Liouvillian is maintained under the mean-field approximation;
(ii) 
because of the detailed balance condition,
the steady state remains the same Gibbs distribution as before the approximation; and
(iii) the particle number is effectively conserved, which is confirmed through numerical simulations.
See Appendix~\ref{Appendix:meanfield} for details.

\subsection{Topological invariants}

The Liouvillian skin effects are induced by nontrivial point gap topology unique to non-Hermitian systems~\cite{Okuma_PRL2020}. The point gap at $\Lambda_0\in\mathbb{C}$ opens when no eigenvalue is equal to $\Lambda_0$.
In the following, we discuss one-dimensional $\mathbb{Z}$ and $\mathbb{Z}_2$ invariants by focusing on a subsystem specified by $k_y$ under periodic boundary conditions (PBC).

As a preparation, by applying a basis transformation (see Appendix~\ref{Appendix:bogoliubov} for details), we rewrite the Liouvillian in a lower triangular form:
\begin{equation}
\label{eq: L triangular}
\mathcal{L} (\vb*{k}) \rightarrow 
\begin{pmatrix}
X^\dagger (\vb*{k}) & 0 \\
* & X(\vb*{k})
\end{pmatrix},
\end{equation}
where the block denoted by the asterisk does not affect the spectrum, namely, the spectrum of the Liouvillian is composed of eigenvalues of the damping matrix $X$~\cite{Song_PRL2019, Liu_PRR2020,Yang_PRR2022,Wang_arXiv2024} and its Hermitian conjugate $X^\dagger$. Therefore, the point gap topology of the Liouvillian is characterized by the matrix $X$ alone. In passing, Eq.~\eqref{eq: L triangular} indicates that the Liouvillian spectrum is symmetric for the real axis.

\subsubsection{Winding number}

For a one-dimensional subsystem specified by $k_y$, the winding number is defined as \cite{wdefineNote, Kawabata_PRX2019}
\begin{equation}
\label{eq:winding}
    w(\Lambda_0) = \frac{1}{2\pi \ii}\int_{k_x=-\pi}^{k_x=\pi} d\ln \det [X(k_x,k_y) - \Lambda_0\1],
\end{equation}
where $\Lambda_0\in \mathbb{C}$ is the reference point. 
The nontrivial point gap topology characterized by a nonzero winding number induces the $\mathbb{Z}$ Liouvillian skin effect~\cite{Okuma_PRL2020}. Specifically, when 
the winding number is positive (negative), the right eigenstates become localized at the left (right) boundary in the $x$ direction, whereas the left eigenstates localize at the opposite boundary~\cite{localizationNote}.

\subsubsection{\texorpdfstring{$\mathbb{Z}_2$}{Z2} invariant}

In the presence of transposed time-reversal symmetry, the system satisfies the following relation:
\begin{equation}
\label{eq: T x T =X}
\mathcal{T}X(k_x, k_y)^{T}\mathcal{T}^{-1} = X(-k_x, -k_y),\quad \mathcal{T} = i\sigma_y,
\end{equation} 
In this case, at $k_y=0$ or $\pi$, a $\mathbb{Z}_2$ topological index $\nu \in {0,1}$ is defined as follows~\cite{Kawabata_PRX2019}:

\begin{equation}
\begin{split}
  & (-1)^{\nu(\Lambda_0)} = \mathrm{sgn}\Bigg[\frac{\mathrm{Pf}\qty[\qty(X(k_x=\pi)-\Lambda_0\1)\mathcal{T}]}{\mathrm{Pf}\qty[\qty(X(k_x=0)-\Lambda_0\1)\mathcal{T}] }\\
  &\times \exp{-\frac{1}{2}\int_{k_x=0}^{k_x=\pi}d\ln\det\qty[\qty(X(k_x)-\Lambda_0\1)\mathcal{T}]
    }
    \Bigg],
\end{split}
\end{equation}
where $\mathrm{Pf}$ denotes Pfaffian.

When 
the $\mathbb{Z}_2$ index takes 1,
Kramers pairs are localized at opposite ends of the system due to time-reversal symmetry~\cite{Okuma_PRL2020}.

\begin{table}[tbp]
\centering
\caption{
Symmetry constraints on the topological invariants $w$ and $\nu$.
$\bigcirc$ and $\times$ indicate the presence and absence of symmetry, respectively. 
“--” indicates that the symmetry may or may not be present, but is not essential in this context.
}
\label{table:topo-index}
\begin{ruledtabular}
\begin{tabular}{@{\hskip 1pt}cccccll@{\hskip 0.5pt}}
$\mathcal{I}M_y$ & $M_x$ & $\mathcal{T}M_y$ & $\mathcal{T}$ & Parameters & $w$ & $\nu$ \\
\hline
$\bigcirc$ & -- & -- & -- & $\alpha = 0$ & 0 & 0 or undef. \\
$\times$   & $\bigcirc$ & $\bigcirc$ & $\times$ & $\alpha \ne 0$, $H_y = 0$, $k_y \ne 0,\pi$ & 0 & undef. \\
$\times$   & $\bigcirc$ & $\bigcirc$ & $\bigcirc$ & $\alpha \ne 0$, $H_y = 0$, $k_y = 0,\pi$ & 0 & 1 \\
$\times$   & $\times$ & $\times$ & $\times$ & $\alpha \ne 0$, $H_y \ne 0$ & $\ne 0$ & undef. \\
\end{tabular}
\end{ruledtabular}
\end{table}

\section{Results\label{Section:Results}}

\subsection{Liouvillian skin effect}

We demonstrate the emergence of $\mathbb{Z}$ and $\mathbb{Z}_2$ Liouvillian skin effects in our electron systems, by numerically diagonalizing the Liouvillian and analyzing symmetry constraints.
The Liouvillian skin effects become pronounced at temperatures below the energy scale of band splitting. 
In addition, the $\mathbb{Z}$ Liouvillian skin effect in our system exhibits scale-free localization of the eigenstates.

\subsubsection{Symmetry constraints}
\label{sec: symm const}

We start with symmetry constraints on the winding number.

The winding number vanishes for arbitrary $\Lambda\in\mathbb{C}$ (see Table~\ref{table:topo-index}) when any of the following symmetry constraints is satisfied~\cite{windingNote}:

{
\def\theenumi{\roman{enumi}}
\def\labelenumi{(\theenumi)}
\begin{enumerate}
\item 
Combined symmetry of inversion $\mathcal{I}$ and mirror reflection $M_y$ with respect to the $y$-axis,  
\begin{eqnarray}
\label{eq: symm_IMy}
\mathcal{I}M_yX(k_x,k_y)(\mathcal{I}M_y)^{-1} &=& X(-k_x,k_y),\\
\mathcal{I} = \sigma_0, \ M_y &=& \ii\sigma_y. \nonumber
\end{eqnarray}
\item
Mirror symmetry $M_x$ in the $x$ direction
\begin{eqnarray}
\label{eq: symm_Mx}
M_{x} X(k_x, k_y) M_{x}^{-1} &=& X(-k_x, k_y),\\ 
M_x &=& \ii\sigma_x. \nonumber
\end{eqnarray}
\item
Combined symmetry of transposed time-reversal operation $\mathcal{T}$ and mirror reflection in the $y$ direction:
\begin{eqnarray}
\label{eq: symm_TMy}
\mathcal{T}M_y X(k_x, k_y)^{T} (\mathcal{T}M_y)^{-1} &=& X(-k_x, k_y),\\ 
\mathcal{T} = M_y &=& \ii\sigma_y. \nonumber 
\end{eqnarray}
\end{enumerate}
}
The constraint in Eq.~\eqref{eq: symm_IMy} forbids the Rashba SOC but allows the Zeeman term $H_y\sigma_y$.
The constraint in Eqs.~\eqref{eq: symm_Mx} or \eqref{eq: symm_TMy} allows the Rashba SOC but forbids the Zeeman term $H_y\sigma_y$ (see Table~\ref{table:topo-index}).
Symmetry constraints on the other types of antisymmetric SOC are discussed in Appendix~\ref{Appendix:otherSOC}.

\subsubsection{Localization of the eigenstates}
\label{sec: localization}

\begin{figure}
\includegraphics[width=0.48\textwidth]{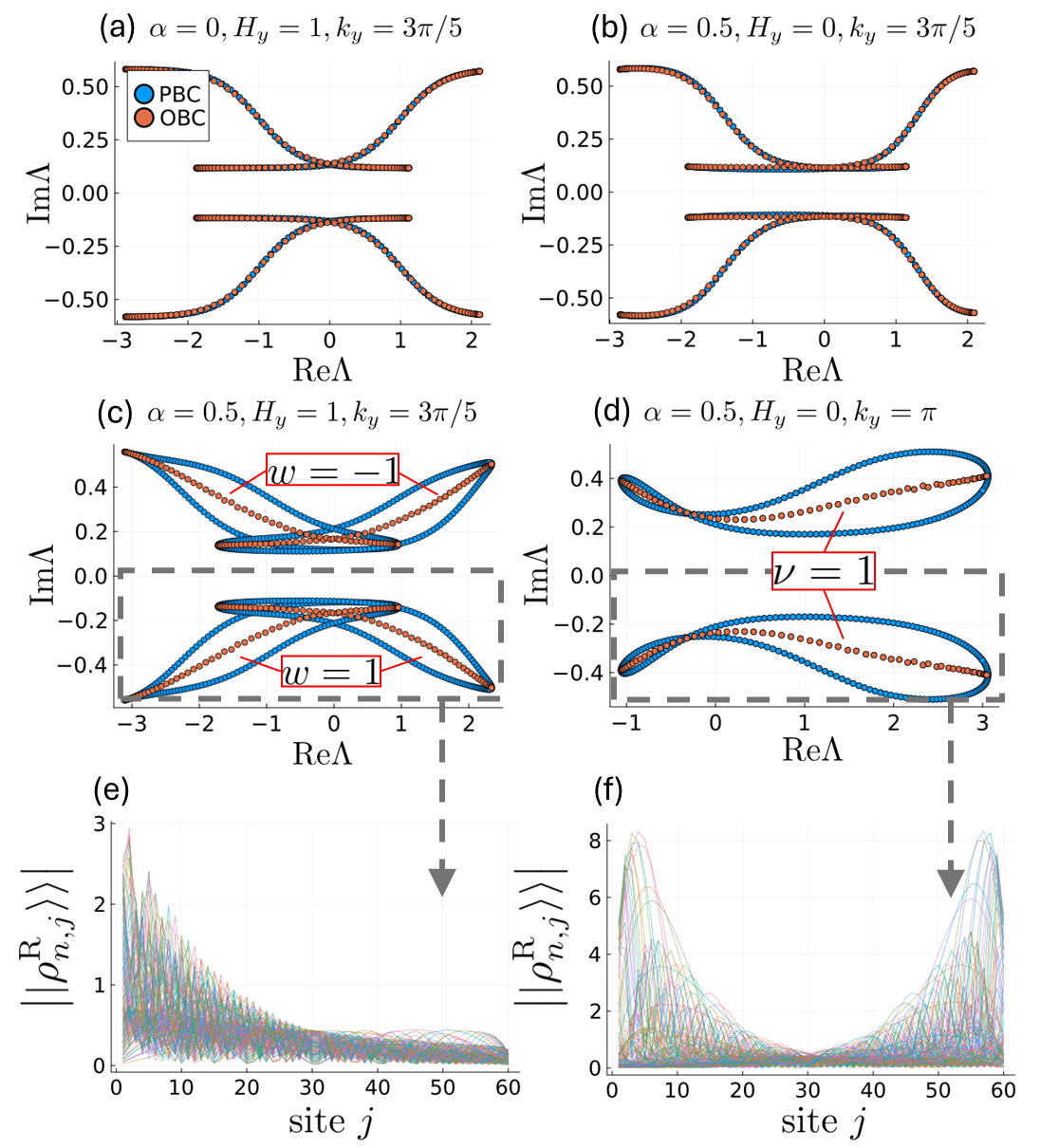}
\caption{
 (a)-(d): Eigenvalues of Liouvillian under OBC or PBC in the $x$ direction. Panel (a) [(b)] displays eigenvalues for ($\alpha$,$H_y$)=(0,1) [($\alpha$,$H_y$)=(0.5, 0)] and $k_y=3\pi/5$, indicating the absence of the Liouvillian skin effect.
Panel (c) [(d)] displays eigenvalues for ($\alpha$,$H_y$)=(0.5,1) and $k_y=3\pi/5$ [($\alpha$,$H_y$)=(0.5,$10^{-6}$)~\cite{Hy0Note} and $k_y=\pi$].
(e) [(f)]: The weight of the right eigenstates under OBC in the $x$ direction for the same parameter set as panel (c) [(d)].
Here, the weight is defined as $\abs{|\rho^{\mathrm{R}}_{n,j}\rangle\rangle} \equiv \sum_{\sigma = \uparrow,\downarrow;\, \tau = \mathrm{K,B}} \abs{|\rho^{\mathrm{R}}_{n,j\sigma\tau}\rangle\rangle}$. The extreme sensitivity of eigenvalues and the localization of eigenstates displayed in panels (c) and (e) [(d) and (f)] are attributed to the nontrivial value of the winding number $w=1$ [$\mathbb{Z}_2$ invariant $\nu=1$].
The data are obtained for $t_{\mathrm{h}}=1$, $\mu=1$,$H_x = 0$, $\gamma=1$, $\gamma_{\mathrm{d}}=0.4$, $T=0.5$, and $L=60$.}

\label{Figure:localization}
\end{figure}

Here, we numerically calculate the Liouvillian spectrum and eigenstates for fixed $k_y$ under various conditions.
First, we discuss the case where either $\alpha$ or $H_y$ is finite.
In this case, for $k_y \neq 0, \pi$, the point gap closes due to the symmetry constraints.
This fact can be confirmed by numerically diagonalizing the Liouvillian [see Figs.~\ref{Figure:localization}(a) and \ref{Figure:localization}(b)].
In these figures, the spectra under PBC coincide with those of open boundary conditions (OBC).
Therefore, the Liouvillian skin effect is absent.

When both $\alpha$ and $H_y$ are finite, the system exhibits the $\mathbb{Z}$ Liouvillian skin effect.
Figure~\ref{Figure:localization}(c) indicates that 
the Liouvillian spectrum under PBC forms loops in the complex plane, acquiring a non zero winding number.
Due to the nontrivial topology with
$w=1$, the loops shrink to lines under OBC. Correspondingly, 
the eigenstates become localized, as shown in Fig~\ref{Figure:localization}(e).
We also note that for $w=-1$, the right eigenstates are localized at the right edge. 
The above observations indicate the emergence of the $\mathbb{Z}$ Liouvillian skin effect.

In passing, reversing the sign of $H_y$ flips the sign of the winding number $w$ for a given $k_y$.
This fact can be seen in Eq.~\eqref{Equation:band}; the sign flip of $H_y$ reverses the winding direction along $k_x$.
As a result of the sign flip of $w$, the eigenstates become localized at the opposite boundary.

Furthermore, for $\alpha \neq 0$ and $H_y=0$, the system exhibits the $\mathbb{Z}_2$ Liouvillian skin effect at $k_y=0,\pi$ where the Liouvillian satisfies the transposed time-reversal symmetry [see Eq.~\eqref{eq: T x T =X}].
As shown in Fig.~\ref{Figure:localization}(d), the spectrum under PBC forms loops at $k_y=\pi$. For each loop, the $\mathbb{Z}_2$ invariant is $\nu=1$, and these loops shrink to lines under OBC. 
Correspondingly, the eigenstates are localized at both ends of the system [Fig.~\ref{Figure:localization}(f)].
These observations indicate the emergence of the $\mathbb{Z}_2$ Liouvillian skin effect protected by the transposed time-reversal symmetry.

\subsubsection{Temperature dependence}

\begin{figure}[htbp]
\includegraphics[width=0.48\textwidth]{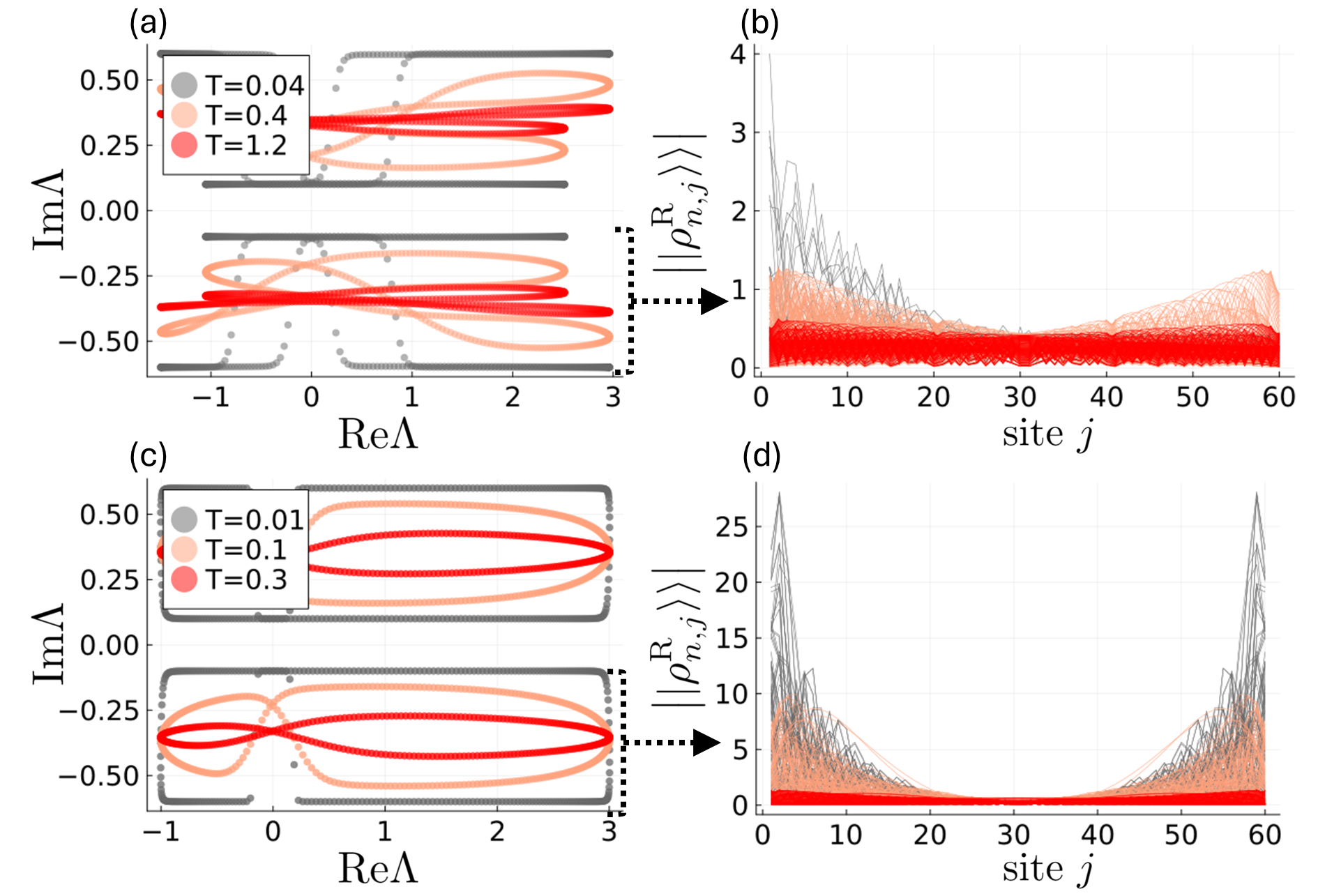}
\caption{\label{Figure:temp}
Temperature dependence of the Liouvillian spectrum and localization of the eigenstates. 
(a) and (c):
Eigenvalues of the Liouvillian under PBC for $T = 0.04$, $0.4$, and $1.2$ in (a), and for $T = 0.01$, $0.1$, and $0.3$ in (c).
Panel (a) [(c)] is obtained for ($\alpha$, $H_y$, $k_y$) = (0.2, 0.4, $5\pi/6$) [(0.1, $10^{-6}$, $\pi$)]~\cite{ParametersNote}.
(b) [(d)]:
The weight of the right eigenstates under OBC in the $x$ direction obtained for the same parameters as those of panel (a) [(c)]. 
All data are obtained for $t_{\mathrm{h}} = 1$, $\mu = 1$, $H_x = 0$, $\gamma = 1$, $\gamma_{\mathrm{d}} = 0.4$, and $L = 60$.
}
\end{figure}

We show that the Liouvillian skin effects become pronounced at temperatures below the energy scale of band splitting (i.e., Rashba SOC or Zeeman splitting).
We begin by considering the low-temperature limit $T \rightarrow 0$, in which the eigenvalues of the Liouvillian given by Eq.~\eqref{Equation:analytic} become
\begin{widetext}
\begin{equation}
\Lambda_{T \rightarrow 0} =
\left\{
\begin{array}{ll}
\begin{array}{l}
\varepsilon_{\vb*{k}1} \pm \dfrac{\ii}{4} \left( \gamma + \gamma_{\mathrm{d}} \right), \
\varepsilon_{\vb*{k}2} \pm \dfrac{\ii}{4} \left( \gamma + \gamma_{\mathrm{d}} \right)
\end{array}
& \text{for } 
\varepsilon_{\vb*{k}1} = \varepsilon_{\vb*{k}2} \\
[8pt]
\begin{array}{l}
\varepsilon_{\vb*{k}1} \pm \dfrac{\ii}{2} \left[ \gamma \, \Theta(\varepsilon_{\vb*{k}2}) + \dfrac{\gamma_{\mathrm{d}}}{2} \right], \
\varepsilon_{\vb*{k}2} \pm \dfrac{\ii}{2} \left[ \gamma \, \Theta(\varepsilon_{\vb*{k}1}) + \dfrac{\gamma_{\mathrm{d}}}{2} \right]
\end{array}
& \text{for } 
\varepsilon_{\vb*{k}1} > \varepsilon_{\vb*{k}2},
\end{array}
\right. 
\end{equation}
\end{widetext}
where $\Theta(x)$ takes $1$, $1/2$, and $0$ for $x>0$, $x=0$, and $x<0$, respectively.
When the SOC and magnetic fields are absent and the energy eigenvalues are degenerate ($\varepsilon_{\vb*{k}1} = \varepsilon_{\vb*{k}2}$), the eigenvalues of Liouvillian are also degenerate, and their imaginary parts become constant. 
As a result, the point gap closes, and the eigenstates do not exhibit localization. In contrast, when the energy eigenvalues split ($\varepsilon_{\vb*{k}1} > \varepsilon_{\vb*{k}2}$), the discontinuity of the step function at the Fermi surface causes a jump in the imaginary part of the Liouvillian spectrum. 
This jump opens the point gap for $T=0.04$ with $(\alpha, H_y, k_y) = (0.2, 0.4, 5\pi/6)$ [Fig.~\ref{Figure:temp}(a)] and for $T=0.01$  
$(\alpha, H_y, k_y) = (0.1, 0, \pi)$ 
[Fig.~\ref{Figure:temp}(c)],  
leading to the localization of the eigenstates of the Liouvillian, as shown in Figs.~\ref{Figure:temp}(b) and~\ref{Figure:temp}(d).

Such localization of eigenstates is suppressed when the temperature exceeds the typical value of $\eta_{\vb*{k}}$. Indeed, by performing a Taylor expansion of the Liouvillian eigenvalues around $\eta_{\vb*{k}}/T = 0$, we obtain
\begin{equation}
    \begin{split}
        \Lambda =&
\varepsilon_{\vb*{k}1} \pm \dfrac{\ii}{4} \left[ \gamma + \gamma_{\mathrm{d}} + \frac{e^{\beta \xi_{\vb*{k}}} -1}{e^{\beta \xi_{\vb*{k}}}+1}\frac{\eta_{\vb*{k}}}{T} + \mathcal{O}\qty(\qty(\frac{\eta_{\vb*{k}}}{T})^2)\right],\\
&\varepsilon_{\vb*{k}2} \pm \dfrac{\ii}{4} \left[ \gamma + \gamma_{\mathrm{d}} - \frac{e^{\beta \xi_{\vb*{k}}} -1}{e^{\beta \xi_{\vb*{k}}}+1}\frac{\eta_{\vb*{k}}}{T} + \mathcal{O}\qty(\qty(\frac{\eta_{\vb*{k}}}{T})^2)\right].
    \end{split}
\end{equation}
Since the imaginary parts of the zeroth-order terms are constant, the point gap closes as the temperature increases beyond the typical value of $\eta_{\vb*{k}}$, leading to the disappearance of localization.

The suppression of the eigenstate localization is also supported by numerical results.
In the case of the $\mathbb{Z}$ Liouvillian skin effect, the eigenstates are localized at $T=0.4$ [see Figs.~\ref{Figure:temp}(a) and~\ref{Figure:temp}(b)], where the energy scale of the band splitting is estimated as $|\eta_{\vb*{k}}| \sim \alpha + \mu_B H_y=0.4$. 
However, as the temperature increases to $T = 1.2$, the region enclosed by the PBC spectrum shrinks significantly, and the localization is strongly suppressed.
A similar tendency is observed for the $\mathbb{Z}_2$ skin effect. 
Figures~\ref{Figure:temp}(c) and~\ref{Figure:temp}(d) indicate that the eigenstates are localized at $T = 0.1$ where the energy scale of the band splitting is estimated as $|\eta_{\vb*{k}}| \sim \alpha =0.1$. However, at $T = 0.3$, the PBC spectrum shrinks significantly and the localization of the eigenstates is strongly suppressed.

\subsubsection{System-size dependence}

\begin{figure}[htbp]
\includegraphics[width=0.48\textwidth]{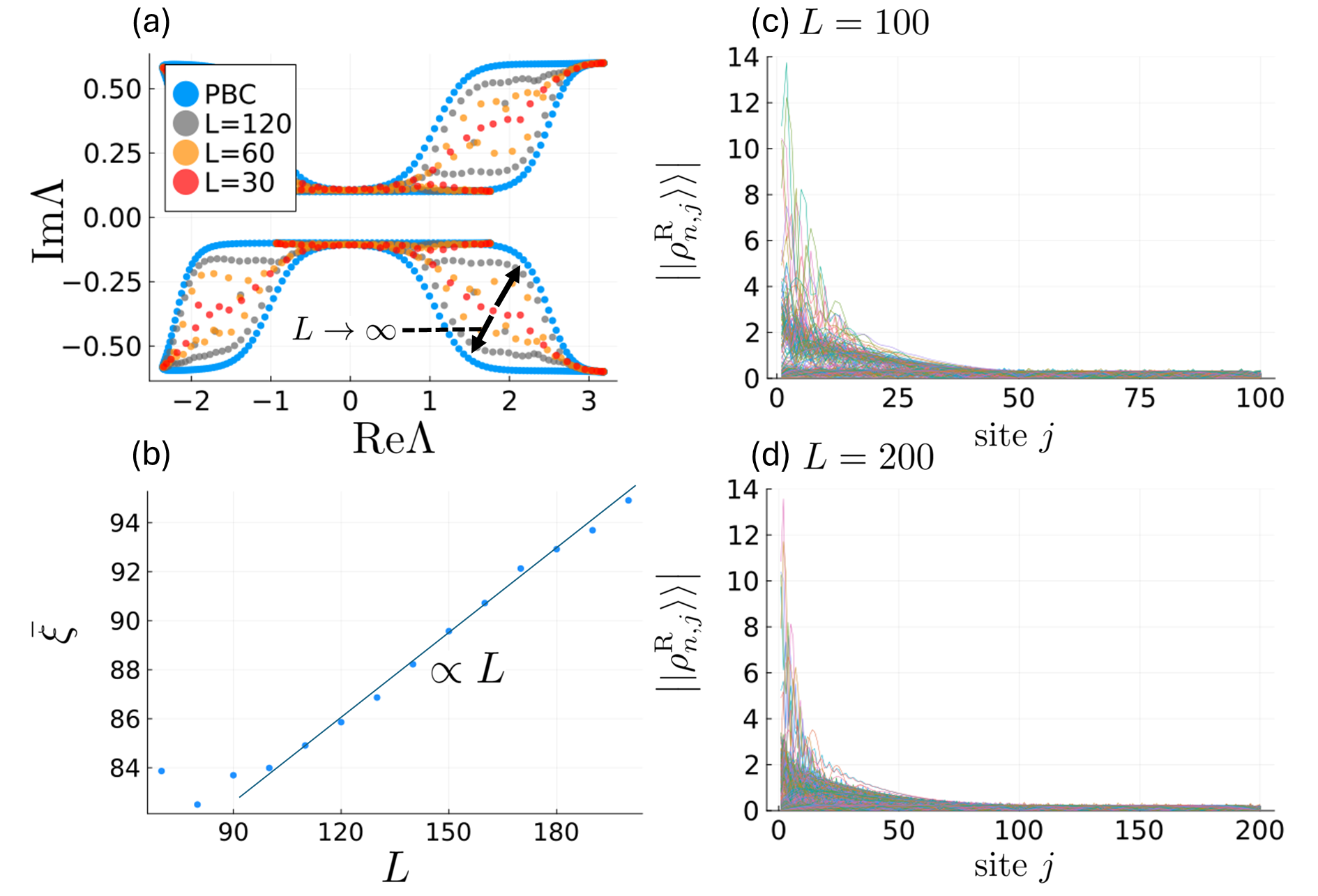}
\caption{System-size dependence of the Liouvillian skin effect.
(a): Eigenvalues of the Liouvillian under OBC for $L = 30$ (red), 60 (orange), and 120 (gray), and under PBC for $L=200$ (blue).
(b): Average localization length $\bar{\xi}$ as a function of the system size $L$ in the $x$ direction. The quantity $\bar{\xi}$ is defined as the average of the localization lengths over all eigenstates. 
(c) and (d):
Spatial profiles of the right eigenstates under OBC for $L = 100$ and $L = 200$, respectively.
The temperature is set to $T = 0.2$ in panels (a), (c), and (d), and $T = 0.3$ in panel (b). Other parameters are fixed as $t_{\mathrm{h}} = 1$, $\mu = 1$, $\alpha = 0.7$, $H_x = 0$, $H_y = 1$, $\gamma = 1$, $\gamma_{\mathrm{d}} = 0.4$, $k_y = 3\pi/4$.
\label{Figure:Size}
}
\end{figure}

We show that the $\mathbb{Z}$ Liouvillian skin effect exhibits system-size dependence, which is a hallmark feature of the non-Hermitian critical skin effect~\cite{Z2Note}. 
Figure~\ref{Figure:Size}(a) indicates that the Liouvillian spectrum under OBC gradually approaches the spectrum under PBC as the system size $L$ in the $x$ direction increases. In addition, Figs.~\ref{Figure:Size}(c) and \ref{Figure:Size}(d) indicate that the eigenstates exhibit scale-free localization. Furthermore, the average localization length $\bar{\xi}$, computed over all eigenstates, increases linearly on $L$ [see Fig.~\ref{Figure:Size}(b)]

The above behaviors are hallmark features of the non-Hermitian critical skin effect ~\cite{Li_NatComm2020,Liu_PRR2020,Yokomizo_PRB2021,Qin_PRB2023,Ji_arXiv2025}.
These behaviors are attributed
to the structure of the quadratic Liouvillian, which takes the same form as that of coupled Hatano–Nelson chains, which is a representative model of the critical skin effect (see Appendix~\ref{Appendix:nhcse}).

In the next subsection, we demonstrate that the system-size dependence of the localization length has a significant impact on the relaxation time.

\subsection{Dynamical properties}
In this subsection, we discuss how the properties of the Liouvillian skin effect influence the dynamics of electrons.

\subsubsection{Particle number imbalance}

\begin{figure*}[htbp]
\includegraphics[width=\textwidth]{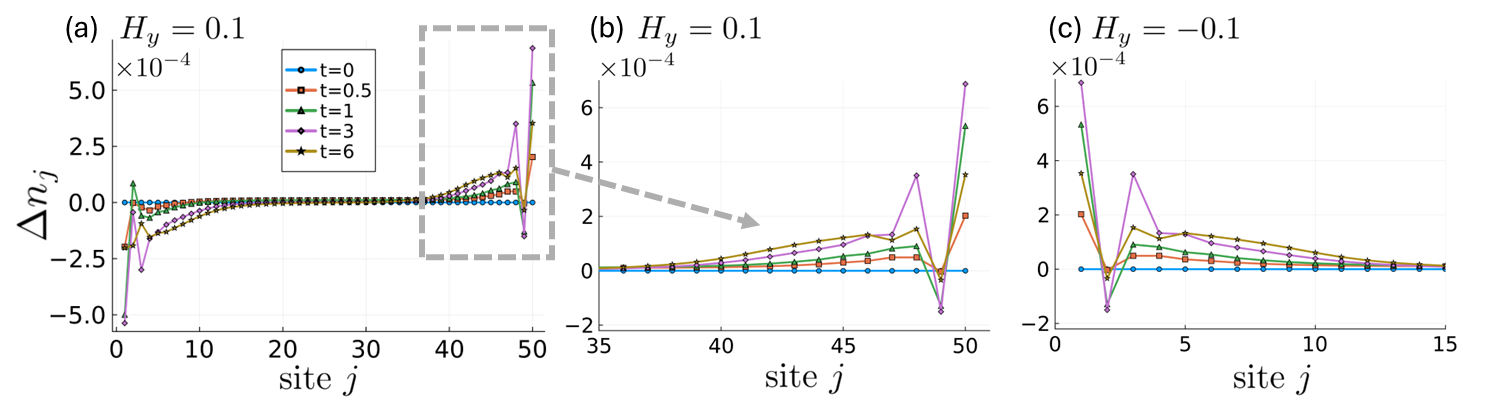}
\caption{\label{Figure:accumulation}
Dynamical accumulation of electrons near the system boundary.
(a): Time evolution of the deviation in electron number $\Delta n_j$ from its initial value, for $t=0$ (circle), $0.5$ (square), $1$ (triangle), $3$ (diamond), and $6$ (star), with $H_y = 0.1$.
(b): Magnified
version of the right-edge region in panel (a).
(c): Electron accumulation near the left edge for opposite magnetic field $H_y = -0.1$.
To ensure that the bulk particle density matches between the initial and steady states, the initial chemical potential was slightly adjusted to $\mu = 0.9998945$, while the time evolution was computed using $\mu = 1$.
All data are obtained for $t_{\mathrm{h}} = 1$, $\alpha=0.5$, $H_x = 0$, $\gamma = 1$, $\gamma_{\mathrm{d}} = 1$, $k_y = 3\pi/4$, $T=0.5$, and $L = 50$.
}
\end{figure*}

We demonstrate that the $\mathbb{Z}$ Liouvillian skin effect leads to dynamical electron accumulation under quench dynamics, where a magnetic field $H_y$ is suddenly applied.
As the initial state, we consider the equilibrium state of the system with SOC but no magnetic field.

First, we obtain the time evolution of the expectation value from eigenvalues and eigenstates of the Liouvillian.
We denote the right [left]  
eigenstates of the Liouvillian by $\phi_{n,(i\sigma\tau)}^{\mathrm{R}}$ 
[$\phi_{n,(i\sigma\tau)}^{\mathrm{L}}$], where $\tau = \mathrm{K}, \mathrm{B}$ and the corresponding eigenvalue is $\Lambda_n$. Using these quantities, the time evolution of the electron number is given by
\begin{equation}
\begin{split}
   & \langle n_i(t)\rangle=\sum_{\sigma,n>0}\phi _{n,(i\sigma \mathrm{K})}^{\mathrm{L}}\phi_{n,(i\sigma \mathrm{K})}^{\mathrm{R}}\\
       & +
        \sum_{n>0,m<0,\sigma}
        e^{-\ii(-\Lambda_n+\Lambda_m)t}
        a_{nm}\phi _{n,(i\sigma \mathrm{K})}^{\mathrm{L}}\phi_{m,(i\sigma \mathrm{K})}^{\mathrm{R}}
        \label{Equation:nt}
\end{split}
\end{equation}
(see Appendix~\ref{Appendix:formula} for details). Here, $n > 0$ ($m < 0$) labels eigenvalues with the positive (negative) imaginary parts. 
The first term represents the electron number in the steady state $(t\rightarrow\infty)$. The second term describes transient dynamics, with the overlap coefficient $a_{nm}$ given by
\begin{equation} a_{nm}=\sum_{\substack{n'>0,i',j'\\\sigma_1,\sigma_2,\tau,\tau'}}\phi_{n,(i'\sigma_1 \tau)}^{\mathrm{R}}\phi_{m,(j'\sigma_2 \tau')}^{\mathrm{L}}\phi_{n',(i'\sigma_1\tau)}^{0\mathrm{L}}\phi_{n',(j'\sigma_2\tau')}^{0\mathrm{R}},
\end{equation}
where $\phi^{0\mathrm{R}}$ ($\phi^{0\mathrm{L}}$) denotes the right (left) eigenstates of the Liouvillian corresponding to the initial state.

The change in the electron number from its initial value is given by
\begin{equation}
\begin{split}
  &\Delta n_i(t) \equiv \langle n_i(t)\rangle - \langle n_i(0)\rangle\\
  &=\sum_{n>0,m<0,\sigma}
        \left\{e^{-\ii(-\Lambda_n+\Lambda_m)t}-1\right\}
        a_{nm}\phi _{n,(i\sigma \mathrm{K})}^{\mathrm{L}}\phi_{m,(i\sigma \mathrm{K})}^{\mathrm{R}}.  
\end{split}
\end{equation}
Since the Liouvillian spectrum is symmetric with respect to the real axis, eigenvalues with positive and negative imaginary parts appear in pairs and always have opposite winding numbers, suggesting the localization of $\phi_{n>0}^{\mathrm{R}}$ and $\phi_{m<0}^{\mathrm{R}}$ at opposite edges. Because the left eigenvectors are localized at the edge opposite that of the right eigenvectors,
$\phi_{n>0}^{\mathrm{L}}$ and $\phi_{m<0}^{\mathrm{R}}$ tend to be localized at the same edge of the system. Therefore, $\Delta n_i(t)$ is expected to be affected by the spatial localization of the eigenstates.
In contrast, for the particle number in the steady state [the first term in Eq.~\eqref{Equation:nt}], the localization effects of $\phi_{n>0}^{\mathrm{L}}$ and $\phi_{n>0}^{\mathrm{R}}$ are mutually canceled, making electron number in the  steady state less sensitive to the localization of eigenstates. These results indicate that the skin effect is manifested more prominently in the transient dynamics than in the steady state. In this paper, we therefore use $\Delta n_i(t)$ to examine the impact of the skin effect.

We numerically compute the time evolution of the electron distribution [see Figs.~\ref{Figure:accumulation}(a) and \ref{Figure:accumulation}(b)] for $k_y=3\pi/4$.
These figures indicate that as time increases, the electron density becomes increasingly unbalanced, leading to electron accumulation at one end of the system.
Moreover, flipping the sign of $H_y$ reverses the direction in which the eigenstates are localized [see Fig.~\ref{Figure:accumulation}(c)], which is due to the sign flip of the winding number $w$.
Besides that, at high temperatures, the dynamical accumulation becomes significantly suppressed.
The above results support that the $\mathbb{Z}$ Liouvillian skin effect leads to the dynamical accumulation of electrons around the boundary.

\subsubsection{Relaxation time}

It is known that when the Liouvillian skin effect occurs, the relaxation time $\tau$ is typically system-size dependent~\cite{Haga_PRL2021,Yang_PRR2022,Hamanaka_PRB2023}. 
Specifically, the relaxation time $\tau$ is estimated from the Liouvillian gap $\Delta$ and the localization length $\xi$ of the slowest relaxing mode~\cite{Haga_PRL2021}:
\begin{equation}
    \tau \sim \frac{1}{\Delta} + \frac{L}{\xi \Delta}.
\end{equation}
In contrast, the relaxation time of our $\mathbb{Z}$ Liouvillian skin effect is independent of the system size $L$ in the thermodynamic limit. This is because the localization length of our Liouvillian skin effect increases linearly on $L$ [see Fig.~\ref{Figure:Size}(b)], namely, using the localization length $\xi = aL + b$ with constants $a$ and $b$, we obtain 
\begin{equation}
    \tau \sim \frac{1}{\Delta} + \frac{L}{(aL + b)\Delta}\ \xrightarrow[L \to \infty]{}\ \sim \frac{1}{\Delta}.
\end{equation}

\begin{figure}[htbp]
\includegraphics[width=0.45\textwidth]{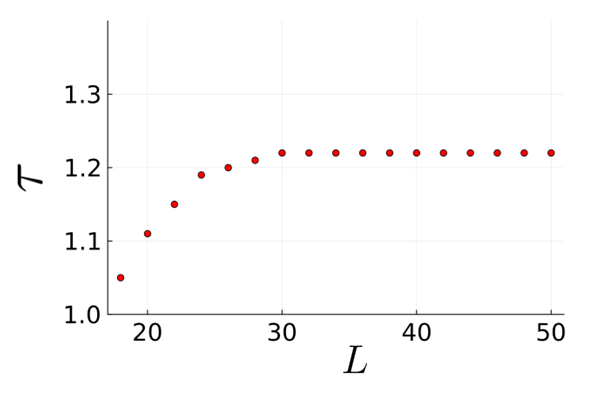}
\caption{\label{Figure:relaxation}
System-size dependence of the relaxation time $\tau$ for $ 18 \leq L \leq 50 $.
The relaxation time saturates as $L$ increases.
The data are obtained for $l_\mathrm{a} = 8$, $t_{\mathrm{h}} = 1$, $\alpha=0.5$, $H_x = 0$, $H_y = 0.1$, $\gamma = 1$, $\gamma_{\mathrm{d}} = 1$, $k_y = 3\pi/4$, and $T = 0.5$.
}
\end{figure}

We numerically verify this behavior by computing the relaxation time $\tau$, which is defined through the relaxation of the average particle number near the boundary~\cite{Haga_PRL2021}:
\begin{equation}
\begin{split}
    &O(t) = \frac{1}{l_{\mathrm{a}}}\sum_{i\le l_{\mathrm{a}}}n_{i}(t),\\
    &|O(\tau)-O(\infty)| = e^{-1}|O(0)-O(\infty)|.
\end{split}
\end{equation}
Here, $l_{\mathrm{a}}$ denotes the number of sites over which the average is taken. As shown in Fig.~\ref{Figure:relaxation}, the numerically obtained relaxation time becomes independent of $L$ 
for $L\gtrsim 30$, which is consistent with the above estimation.

This behavior can be intuitively understood as follows.
The magnetic field induces a particle number imbalance in real space [see Fig.~\ref{Figure:accumulation} (a)]. This imbalance does not require global transport of electrons across the system; it can be achieved by local rearrangements of electrons near each site.
As a result, the relaxation process does not involve transport over distances proportional to the system size $L$, and thus the relaxation time $\tau$ does not scale with $L$.

\section{Conclusion}
\label{Section:Conclusion}

We have proposed a two-dimensional electron system on a substrate as a new platform for Liouvillian skin effects. This system exhibits $\mathbb{Z}$ and $\mathbb{Z}_2$ Liouvillian skin effects due to the interplay among energy dissipations, Rashba SOC, and a transverse magnetic field. In our electron system, the Liouvillian skin effects become pronounced below the energy scale of the band splitting induced by Rashba SOC and the Zeeman splitting; in the high temperature regions, dissipations become homogeneous in momentum space, suppressing the skin effects.
While our $\mathbb{Z}$ Liouvillian skin effect leads to charge accumulation, its relaxation time is independent of the system size, in contrast to previously reported Liouvillian skin effects. This difference is attributed to the scale-free behavior of the localization length analogous to non-Hermitian critical skin effects.

\section*{Acknowledgments}
The authors thank Y. Hirobe, K. Hara, S. Hamanaka, S. Kaneshiro, M. Sato, and M. Sigrist for fruitful discussion.
This work is supported by JSPS KAKENHI Grant Nos.~JP21K13850, JP23KK0247, JP25K07152, and JP25H02136, as well as JSPS Bilateral Program No.~JPJSBP120249925.
T.~Y.~is grateful for the support from the ETH Pauli Center for Theoretical Studies and the Grant from Yamada Science Foundation.

\appendix
\section{General properties of the vectorized formalism\label{Appendix:properties}}

We first summarize some fundamental properties of vectorization [see Eq.~\eqref{Equation:vectorization}], and derive the relations presented in the main text following Refs.~\cite{Shibata_PRB2019,Yoshida_PRR2020,Hanai_arXiv2024}.

The inner product between two vectorized matrices, known as the Hilbert–Schmidt inner product, is expressed as
\begin{equation}
\langle\langle A|B\rangle\rangle = \tr(A^{\dagger}B).
\label{Equation:innerproduct}
\end{equation}
Another important identity is
\begin{equation}
\lvert A \rho B\rangle\rangle = (A\otimes B^{T})\lvert \rho \rangle\rangle,
\label{Equation:vecrelation}
\end{equation}
which can be proved by the following calculation
\begin{equation}
  \begin{split}
    & (A\otimes B ^{T})\lvert \rho \rangle\rangle =(A\otimes B ^{T})\sum_{i', j'} \rho_{i'j'} \lvert \phi _{i'}\rangle\rangle\otimes \lvert \phi _{j'}\rangle\rangle \\
    &= \sum_{i,i',j,j'} A_{ii'}B_{jj'}^{T}\rho_{i'j'} \lvert \phi _{i}\rangle\rangle\otimes \lvert \phi _{j}\rangle\rangle =\lvert A \rho B\rangle\rangle .
  \end{split}
\end{equation}

By making use of  Eqs.~\eqref{Equation:innerproduct} and~\eqref{Equation:vecrelation}, we have
\begin{equation}
    \tr(A\rho B) = \langle\langle\1|A\rho B\rangle\rangle = \langle\langle\1|A\otimes B^{T}|\rho\rangle\rangle,
\end{equation}
which corresponds to Eq.~\eqref{Equation:expectationval}.

Moreover, by Eq.~\eqref{Equation:vecrelation}, we have
\begin{equation}
    \begin{split}
        &\rho H\rightarrow (H^T)|\rho\rangle\rangle=(H^*)|\rho\rangle\rangle,\\
        &L _{lm}\rho L _{lm} ^{\dagger}\rightarrow L _{lm}\otimes L _{lm} ^{*} |\rho\rangle\rangle,\\
        &\rho L_{lm}^\dagger L_{lm}\rightarrow \1\otimes \qty(L _{lm}^{\dagger}L _{lm}) ^{*}.
    \end{split}
\end{equation}
Therefore, the Liouvillian superoperator $\mathscr{L}$ is mapped to a non-Hermitian linear operator $\mathcal{L}$ as shown in Eq.~\eqref{Equation:vecGKSL}.

We assume $\mathcal{L}$ is diagonalizable so that its right and left eigenvectors form a biorthonormal basis:
\begin{equation}
    \langle\langle \rho_{n}^{\mathrm{L}} | \rho_{m}^{\mathrm{R}} \rangle\rangle = \delta_{nm}, \quad 
    \sum_{n} |\rho_{n}^{\mathrm{R}} \rangle\rangle \langle\langle \rho_{n}^{\mathrm{L}} | = \1.
    \label{Equation:biorthonormal}
\end{equation} 
This assumption will be maintained throughout the rest of this paper.
Using this condition, the density matrix at time $t$ can be decomposed in terms of the eigenstates of the Liouvillian as 
\begin{equation}
    |\rho (t)\rangle\rangle = e^{-\ii\mathcal{L}t} |\rho(0)\rangle\rangle 
    = \sum_{n} e^{-\ii\Lambda_{n} t} |\rho _{n}^{\mathrm{R}} \rangle\rangle \langle\langle \rho _{n}^{\mathrm{L}} | \rho(0) \rangle\rangle.
    \label{Equation:expansion}
\end{equation}
This expansion is valid only when the Liouvillian $\mathcal{L}$ is time-independent. 
In the present work, we consider a time-independent Liouvillian given by Eq.~\eqref{Equation:Liouvillian_op}, and thus this decomposition can be applied.

Furthermore, when the Liouvillian preserves trace, as ensured by the structure of the GKSL equation, the following relation holds:
\begin{equation}
    \tr(\mathscr{L}[\rho(t)])
    = \langle\langle \1 | \mathcal{L} |\rho(t)\rangle\rangle = 0
\end{equation}
for any state $|\rho(t)\rangle\rangle$. This implies that
\begin{equation}
    \langle\langle \1 | \mathcal{L} = 0,
    \label{Equation:left0}
\end{equation}
which means that $\langle\langle \1 |$ is the left eigenvector of $\mathcal{L}$ with eigenvalue zero. We will use this fact in Appendix~\ref{Appendix:bogoliubov} and Appendix~\ref{Appendix:formula}.

\section{Bogoliubov quasiparticle picture of the Liouvillian\label{Appendix:bogoliubov}}
As a preliminary step toward deriving expressions for physical observables, we follow the procedure in Ref.~\cite{Wang_arXiv2024} and express the quadratic Liouvillian in terms of ``Bogoliubov quasiparticles"~\cite{BogoliubovNote,Yamamoto_PRL2019,Soma_PRB2024_1,Soma_PRB2024_2}.

Under the mean-field approximation introduced in the main text, the Liouvillian takes the form of a GKSL equation with single-particle gain and loss jump operators,
\begin{equation}
L_{\mu}^{\mathrm{g}} = \sum_{i\sigma} D^{\mathrm{g}}_{\mu, i\sigma } c_{i\sigma}^\dagger,\quad
L_{\mu}^{\mathrm{l}} = \sum_{i\sigma} D^{\mathrm{l}}_{\mu,i\sigma} c_{i\sigma},
\end{equation}
whose derivation is presented in Appendix~\ref{Appendix:meanfield}.
Here, the index $\mu$ labels the individual gain and loss processes, $i$ denotes the site, and $\sigma = \uparrow,\downarrow$ represents the spin.

For such jump operators, the vectorized Liouvillian takes the form
\begin{equation}
\begin{split}
    &\mathcal{L} = \sum_{i,j,\sigma,\sigma'}\Big[\qty(H_{i\sigma,j\sigma'}c_{i\sigma \mathrm{K}}^{\dagger}c_{j\sigma' \mathrm{K}} - H_{i\sigma,j\sigma'}^*c_{i\sigma \mathrm{B}}^{\dagger}c_{j\sigma' \mathrm{B}} ) \\
    &+2\ii\eta\qty(
    M_{i\sigma,j\sigma'}^{\mathrm{g}*}c_{i\sigma\mathrm{K}}^\dagger c_{j\sigma'\mathrm{B}}^\dagger -M_{i\sigma,j\sigma'}^{\mathrm{l}*}c_{i\sigma\mathrm{K}} c_{j\sigma'\mathrm{B}} 
    )\\
    &-\ii\qty(
    M_{i\sigma,j\sigma'}^{\mathrm{g}*}c_{i\sigma\mathrm{B}} c_{j\sigma'\mathrm{B}}^\dagger + M_{i\sigma,j\sigma'}^{\mathrm{g}}c_{i\sigma\mathrm{K}} c_{j\sigma'\mathrm{K}}^\dagger 
    )\\
    &-\ii\qty(
    M_{i\sigma,j\sigma'}^{\mathrm{l}*}c_{i\sigma\mathrm{B}}^\dagger c_{j\sigma'\mathrm{B}} + M_{i\sigma,j\sigma'}^{\mathrm{l}}c_{i\sigma\mathrm{K}}^\dagger c_{j\sigma'\mathrm{K}} 
    )\Big],
    \label{Equation:Liouvillian_vec}
\end{split}
\end{equation}
where the matrices $M_{i\sigma,j\sigma'}^{\mathrm{g(l)}}$ are defined as
\begin{equation}   M_{i\sigma,j\sigma'}^{\mathrm{g(l)}} = \frac{1}{2}\sum_\mu D^{\mathrm{g(l)}*}_{\mu,i\sigma} D^{\mathrm{g(l)}}_{\mu ,j\sigma'},
\end{equation}
and $\eta$ is defined in Eq.~\eqref{Equation:eta}.

Since $\mathcal{L}$ and $\eta$ commute, i.e., $ [\mathcal{L}, \eta] = 0 $, the action of $\mathcal{L}$ can be restricted to the subspace with $\eta = 1$~\cite{Prosen_NJoP2008}. This subspace corresponds to states where the total particle number in the ket and bra spaces is even. In this work, we focus on physical observables composed of an even number of fermions, such as the particle number, so this projection does not affect the analysis.
The projected Liouvillian can then be represented in the Nambu basis
\begin{equation}
\vb{c}=
\begin{pmatrix}
c_{1\sigma\mathrm{K}}&\dots& c_{L\sigma\mathrm{K}}&c_{1\sigma\mathrm{B}}^\dagger&\dots& c_{L\sigma\mathrm{B}}^\dagger
\end{pmatrix},
\end{equation}
as follows:
\begin{equation}
    \begin{split}
        \mathcal{L}|_{\eta=1} &= \vb{c}^\dagger \vb{L} \vb{c} 
        - \mathrm{tr}\big(\ii M^{\mathrm{l}} + \ii (M^{\mathrm{g}})^T + H\big), \\
        \vb{L} &= 
        \begin{pmatrix}
            H + \ii (M^{\mathrm{g}})^T - \ii M^{\mathrm{l}} & 2\ii (M^{\mathrm{g}})^T \\
            2\ii M^{\mathrm{l}} & H - \ii (M^{\mathrm{g}})^T + \ii M^{\mathrm{l}}
        \end{pmatrix},
        \label{Equation:Lmatrix}
    \end{split}
\end{equation}
where we have used the relations $H_{i\sigma,j\sigma'}^* = H_{i\sigma,j\sigma'}^T$ and $M_{i\sigma,j\sigma'}^* = M_{i\sigma,j\sigma'}^T$.

Next, we discuss the spectrum of the Liouvillian, which is essential for defining Bogoliubov quasiparticles.
We apply the unitary transformation
\begin{equation}
    H = \frac{1}{\sqrt{2}}
    \begin{pmatrix}
        \1 & \1 \\
        \1 & -\1
    \end{pmatrix},
\end{equation}
under which the matrix $\vb{L}$ is brought into a lower triangular form:
\begin{equation}
    H \vb{L} H^{-1} = 
    \begin{pmatrix}
        X^\dagger & 0 \\
        2\ii \big[ (M^{\mathrm{g}})^T - M^{\mathrm{l}} \big] & X
    \end{pmatrix},
\end{equation}
where $X = H - \ii (M^{\mathrm{g}})^T - \ii M^{\mathrm{l}}$. Since the spectrum of the Liouvillian is determined by the eigenvalues of $X$ and $X^\dagger$, it is symmetric with respect to the real axis. Furthermore, since both $M^{\mathrm{g}}$ and $M^{\mathrm{l}}$ are positive semi-definite by construction~\cite{positivesemiNote}, the spectrum of $X$ lie in the lower half of the complex plane, while that of $X^\dagger$ lie in the upper half.

Finally, we introduce the Bogoliubov quasiparticles.  
We denote the right [left] eigenvector of $\vb{L}$ by $\phi_{n,(i\sigma\tau)}^{\mathrm{R}}$ [$\phi_{n,(i\sigma\tau)}^{\mathrm{L}}$], where $\tau = \mathrm{K}, \mathrm{B}$, and the corresponding eigenvalue is $\Lambda_n$. Using these eigenvectors, the Bogoliubov quasiparticles are defined as
\begin{equation}
    \begin{split}
    &b_{n \le 0} = \sum_{i\sigma}( \phi_{n \le 0,(i\sigma\mathrm{K})}^{\mathrm{L}} c_{i\sigma\mathrm{K}} + \phi_{n \le 0,(i\sigma\mathrm{B})}^{\mathrm{L}} c_{i\sigma\mathrm{B}}^\dagger), \\
    &\bar{b}_{n \ge 0} = \sum_{i\sigma} (\phi_{n \ge 0,(i\sigma\mathrm{K})}^{\mathrm{L}} c_{i\sigma\mathrm{K}} + \phi_{n \ge 0,(i\sigma\mathrm{B})}^{\mathrm{L}} c_{i\sigma\mathrm{B}}^\dagger ), \\ 
    &\bar{b}_{n \le 0} = \sum_{i\sigma} (\phi_{n \le 0,(i\sigma\mathrm{K})}^{\mathrm{R}} c_{i\sigma\mathrm{K}}^\dagger + \phi_{n \le 0,(i\sigma\mathrm{B})}^{\mathrm{R}} c_{i\sigma\mathrm{B}} ),\\
    &b_{n \ge 0} = \sum_{i\sigma}( \phi_{n \ge 0,(i\sigma\mathrm{K})}^{\mathrm{R}} c_{i\sigma\mathrm{K}}^\dagger + \phi_{n \ge 0,(i\sigma\mathrm{B})}^{\mathrm{R}} c_{i\sigma\mathrm{B}}).
    \label{Equation:bogo}
    \end{split}
\end{equation}
Here, $n\ge0$ ($n\le0$) labels the eigenvalues of $X$ ($X^\dagger$) whose imaginary part is non-negative (non-positive). 
From the biorthonormality condition given in Eq~\eqref{Equation:biorthonormal}, these quasiparticle operators satisfy
the following anticommutation relations:
\begin{equation}
    \{b_n,\;\bar{b}_m\} =\delta_{nm},
 \quad
 \{b_n,\,b_m\} =0,
 \quad
 \{\bar{b}_n,\,\bar{b}_m\} =0.
 \label{Equation:bogo_anticomm}
\end{equation}

Furthermore, using Eq.~\eqref{Equation:biorthonormal}, the inverse transformation can be obtained as
\begin{equation}
    \begin{split}
        &c_{i\sigma \mathrm{K}} = \sum_{n}(\phi_{n\le0,(i\sigma \mathrm{K})}^{\mathrm{R}}b_{n\le0} + \phi_{n\ge0,(i\sigma \mathrm{K})}^{\mathrm{R}}\bar{b}_{n\ge0}),\\
        &c_{i\sigma \mathrm{B}}^\dagger = \sum_{n}(\phi_{n\le0,(i\sigma \mathrm{B})}^{\mathrm{R}}b_{n\le0} + \phi_{n\ge0,(i\sigma \mathrm{B})}^{\mathrm{R}}\bar{b}_{n\ge0}),\\
        &c_{i\sigma \mathrm{K}}^\dagger = \sum_{n}(\phi_{n\le0,(i\sigma \mathrm{K})}^{\mathrm{L}}\bar{b}_{n\le0} + \phi_{n\ge0,(i\sigma \mathrm{K})}^{\mathrm{L}}b_{n\ge0}),\\
        &c_{i\sigma \mathrm{B}} = \sum_{n}(\phi_{n\le0,(i\sigma \mathrm{B})}^{\mathrm{L}}\bar{b}_{n\le0} + \phi_{n\ge0,(i\sigma \mathrm{B})}^{\mathrm{L}}b_{n\ge0}).
        \label{Equation:inversbogo}
    \end{split}
\end{equation}
Using Eqs.~\eqref{Equation:biorthonormal} and~\eqref{Equation:inversbogo}, one can show that
\begin{equation}
    \vb{c}^\dagger \vb{L} \vb{c}  = \sum_{n \ge 0} \Lambda_{n} b_{n} \bar{b}_{n} 
        + \sum_{n \le 0} \Lambda_{n} \bar{b}_{n} b_{n}.
        \label{Equation:cLc}
\end{equation}
Then, using Eqs.~\eqref{Equation:bogo_anticomm} and~\eqref{Equation:cLc}, the Liouvillian can be written as
\begin{equation}
    \begin{split}
        \mathcal{L}|_{\eta=1} &= \sum_{n \ge 0} \Lambda_n b_n \bar{b}_n 
        + \sum_{n \le 0} \Lambda_n \bar{b}_n b_n\\
        &- \tr\big(\ii M^{\mathrm{l}} + \ii (M^{\mathrm{g}})^T + H\big) \\
        &= \sum_{n \ge 0} (-\Lambda_n \bar{b}_n b_n) 
        + \sum_{n \le 0} \Lambda_n \bar{b}_n b_n.
        \label{Equation:diagonalized}
    \end{split}
\end{equation}
Here, we have used the fact that the sum $\sum_{n \ge 0} \Lambda_n$ corresponds to the sum of the eigenvalues of the matrix $X^\dagger$, which cancels the constant term originally present.

When the spectrum of the Liouvillian contains no eigenvalue with zero imaginary part, then Eq.~\eqref{Equation:diagonalized} implies that the right steady state $|\mathrm{SS}\rangle\rangle$ and the left steady state $\langle\langle \mathrm{SS}|$ of the Liouvillian are uniquely defined by
\begin{equation}
    b_n|\mathrm{SS}\rangle\rangle = 0, \quad \langle\langle \mathrm{SS}|\bar{b}_{n}=0 ,\quad \text{for all } n.
    \label{Equation:ss}
\end{equation}
In addition, Eq.~\eqref{Equation:left0} gives
\begin{equation}
    \langle\langle \mathrm{SS}| = \langle\langle \1|.
    \label{Equation:ss1}
\end{equation}
Furthermore, the Liouvillian gap $\Delta$ corresponds to the smallest absolute value of the imaginary part in the spectrum of the Bogoliubov quasiparticle excitations.

\section{Formula for computing physical observables in quench dynamics\label{Appendix:formula}}

We consider quench dynamics in which the system, initially prepared in the steady state of a Liouvillian $\mathcal{L}_{0}|_{\eta=1}$, undergoes time evolution governed by a different Liouvillian $\mathcal{L}|_{\eta=1}$ for $t > 0$ following a sudden quench at $t = 0$.
Here, the Liouvillian is projected onto the $\eta=1$ sector (see Appendix~\ref{Appendix:bogoliubov}), and the Liouvillian of this sector is denoted by $\mathcal{L}|_{\eta=1}$.
To analyze the time evolution, we compute time-dependent physical observables using the Bogoliubov quasiparticles [see Eq.~\eqref{Equation:expt}].  

We focus on the case where the spectra of both $\mathcal{L}_0|_{\eta=1}$ and $\mathcal{L}|_{\eta=1}$ have no eigenvalues with zero imaginary part, so that the initial state and steady state ($t\rightarrow\infty$) are uniquely determined by Eq.~\eqref{Equation:ss}.  
This condition is satisfied in the model considered in the main text when $\gamma_\mathrm{d} > 0$, as confirmed by the analytical expression of the spectrum in Eq.~\eqref{Equation:analytic}.

In this case, due to the biorthonormality condition given in Eq.~\eqref{Equation:biorthonormal}, only the steady states have finite overlap with $\langle\langle \1|$.
Let $|\rho_{\{n\}}^{\mathrm{R}}\rangle\rangle$ and $|\rho_{\{n'\}}^{0\mathrm{R}}\rangle\rangle$ denote the many-body right eigenstates of $\mathcal{L}|_{\eta=1}$ and $\mathcal{L}_0|_{\eta=1}$, respectively,  
where $\{n\}$ and $\{n'\}$ label occupation configurations of the Bogoliubov quasiparticles $\bar{b}_n$ and $\bar{b}_{n'}^0$ defined with respect to $\mathcal{L}|_{\eta=1}$ and $\mathcal{L}_0|_{\eta=1}$.  
Then, the only state that survives the projection by $\langle\langle \1|$ is the vacuum of each corresponding Bogoliubov basis:
\begin{eqnarray}
    \label{Equation:ssortho}
    &\langle\langle \1 | \rho_{\{n\}}^{\mathrm{R}} \rangle\rangle = 
    \begin{cases}
    1 & \text{for } \{n\} = \{0\} \ (\text{steady state}), \\
    0 & \text{otherwise},
    \end{cases} \\
    &\langle\langle \1 | \rho_{\{n'\}}^{0\mathrm{R}} \rangle\rangle = 
    \begin{cases}
    1 & \text{for } \{n'\} = \{0\} \ (\text{initial state}), \\
    0 & \text{otherwise}.
    \end{cases}
    \label{Equation:ssortho_0}
\end{eqnarray}

We then calculate the expectation value of a quadratic operator, $\langle c_{i\sigma}^\dagger c_{j\sigma'} \rangle(t)$, from which physical observables can be constructed. 
Using Eqs.~\eqref{Equation:expectationval} and~\eqref{Equation:expansion}, we obtain
\begin{equation}
\begin{split}
    &\langle c_{i\sigma}^\dagger c_{j\sigma'} \rangle(t) = \tr\big[c_{i\sigma}^\dagger c_{j\sigma'}  \rho(t)\big] 
    = \langle\langle \1|  c_{i\sigma\mathrm{K}}^\dagger c_{j\sigma'\mathrm{K}} |\rho(t)\rangle\rangle \\
    &= \sum_{\{n\}} e^{-\ii\Lambda_{\{n\}} t} 
    \langle\langle \1|c_{i\sigma\mathrm{K}}^\dagger c_{j\sigma'\mathrm{K}}  | \rho_{\{n\}}^{\mathrm{R}} \rangle\rangle 
    \langle\langle \rho_{\{n\}}^{\mathrm{L}} | \rho(0)\rangle\rangle.
\end{split}
\end{equation}
The quantity $\langle\langle \1|c_{i\sigma\mathrm{K}}^\dagger c_{j\sigma'\mathrm{K}}  | \rho_{\{n\}}^{\mathrm{R}} \rangle\rangle$ can be expressed in terms of the Bogoliubov quasiparticles of $\mathcal{L}|_{\eta=1}$ using Eq.~\eqref{Equation:inversbogo} as
\begin{equation}
    \begin{split}
        &\langle\langle \1|c_{i\sigma\mathrm{K}}^\dagger c_{j\sigma'\mathrm{K}}  | \rho_{\{n\}}^{\mathrm{R}} \rangle\rangle \\
        &=\sum_{n<0,m>0} \phi^{\mathrm{L}}_{n,(i\sigma\mathrm{K})}\phi^{\mathrm{R}}_{m,(j\sigma'\mathrm{K})}\langle\langle 1|\bar{b}_{n}\bar{b}_{m} |\rho_{\{n\}}^R\rangle\rangle\\
        &+\sum_{n<0,m<0} \phi^{\mathrm{L}}_{n,(i\sigma\mathrm{K})}\phi^{\mathrm{R}}_{m,(j\sigma'\mathrm{K})}\langle\langle 1|\bar{b}_{n}b_{m} |\rho_{\{n\}}^R\rangle\rangle\\
        &+\sum_{n>0,m>0} \phi^{\mathrm{L}}_{n,(i\sigma\mathrm{K})}\phi^{\mathrm{R}}_{m,(j\sigma'\mathrm{K})}\langle\langle 1|b_{n}\bar{b}_{m} |\rho_{\{n\}}^R\rangle\rangle\\
        &+\sum_{n>0,m<0} \phi^{\mathrm{L}}_{n,(i\sigma\mathrm{K})}\phi^{\mathrm{R}}_{m,(j\sigma'\mathrm{K})} \langle\langle 1|b_{n}b_{m} |\rho_{\{n\}}^R\rangle\rangle,
        \label{Equation:bogo_expansion}
    \end{split}
\end{equation}
where we have replaced $n \ge 0$ ($n \le 0$) with $n > 0$ ($n < 0$) since the spectrum has no eigenvalues with zero imaginary part.
We then evaluate each term on the right-hand side of Eq.~\eqref{Equation:bogo_expansion}.  
The first and second terms vanish due to Eqs.~\eqref{Equation:ss} and~\eqref{Equation:ss1}.  
For the third term, using Eqs.~\eqref{Equation:bogo_anticomm} and~\eqref{Equation:ssortho}, we find
\begin{equation}
    \langle\langle \1 | b_{n>0} \bar{b}_{m>0} | \rho_{\{n\}}^{\mathrm{R}} \rangle\rangle = \delta_{nm} \, \delta_{\{n\},\{0\}}.
\end{equation}
Since $\Lambda_{\{0\}} = 0$ and $\langle\langle \1|\rho(0)\rangle\rangle = 1$, the contribution to $\langle c_{i\sigma}^\dagger c_{j\sigma'} \rangle(t)$ from the third term becomes
\begin{equation}
    \sum_{n>0} \phi^{\mathrm{L}}_{n,(i\sigma\mathrm{K})} \phi^{\mathrm{R}}_{n,(j\sigma'\mathrm{K})}.
    \label{Equation:third}
\end{equation}
This contribution is time-independent and represents the steady state part of the expectation value.
Next, we evaluate the contribution of the fourth term to the expectation value.  
For $\langle\langle \1|b_{n>0} b_{m<0} | \rho_{\{n\}}^{\mathrm{R}} \rangle\rangle$ to be nonzero, the state $|\rho_{\{n\}}^{\mathrm{R}} \rangle\rangle$ must be of the form $\bar{b}_{m<0} \bar{b}_{n>0} |\mathrm{SS}\rangle\rangle$.    
From Eq.~\eqref{Equation:diagonalized}, the corresponding eigenvalue is given by $\Lambda_{\{n>0,m<0\}} = -\Lambda_{n>0} + \Lambda_{m<0}$.  
The overlap with the initial state is computed as
\begin{equation}
    \begin{split}
        &a_{nm}\equiv\langle\langle \rho_{\{n>0,m<0\}}^{\mathrm{L}} | \rho(0)\rangle\rangle = \langle\langle\1 |b_{n>0}b_{m<0}| \rho(0)\rangle\rangle\\
        &=\sum_{\substack{n'>0,i',j'\\\sigma_1,\sigma_2,\tau,\tau'}}\phi_{n>0,(i'\sigma_1 \tau)}^{\mathrm{R}}\phi_{m<0,(j'\sigma_2 \tau')}^{\mathrm{L}}\phi_{n',(i'\sigma_1\tau)}^{0\mathrm{L}}\phi_{n',(j'\sigma_2\tau')}^{0\mathrm{R}},
        \label{Equation:anm}
    \end{split}
\end{equation}
 where $\phi^{0\mathrm{R}}_{i\sigma\tau}$ [$\phi^{0\mathrm{L}}_{i\sigma\tau}$] denotes the right [left] eigenvector of $\mathcal{L}_0|_{\eta=1}$ represented in the Nambu basis.
Here, Eq.~\eqref{Equation:anm} is obtained as follows:
We first rewrite the product of Bogoliubov operators $b_{n>0} b_{m<0}$ in terms of the original fermionic operators $c$ and $c^\dagger$, using Eq.~\eqref{Equation:bogo}.  
We then express these operators in terms of the Bogoliubov operators $b^0$ and $\bar{b}^0$ by applying the inverse transformation Eq.~\eqref{Equation:inversbogo}.  
Finally, we use the condition given in Eq.~\eqref{Equation:ssortho_0} to extract the nonvanishing contributions.
As a result, the contribution of the fourth term to the expectation value is given by
\begin{equation}
    \sum_{n>0,\, m<0} 
    e^{-\ii(-\Lambda_n + \Lambda_m)t}a_{nm}  
    \phi^{\mathrm{L}}_{n,(i\sigma\mathrm{K})} 
    \phi^{\mathrm{R}}_{m,(j\sigma'\mathrm{K})}.
    \label{Equation:fourth}
\end{equation}

Combining this with the steady state contribution in Eq.~\eqref{Equation:third}, the full expression for the expectation value becomes

\begin{equation}
\begin{split}
    &\langle c_{i\sigma}^\dagger c_{j\sigma'} \rangle(t) 
    = \sum_{n>0} 
    \phi^{\mathrm{L}}_{n,(i\sigma\mathrm{K})} 
    \phi^{\mathrm{R}}_{n,(j\sigma'\mathrm{K})} \\
    &+ \sum_{n>0,\, m<0} 
     e^{-\ii(-\Lambda_n + \Lambda_m)t} a_{nm} 
    \phi^{\mathrm{L}}_{n,(i\sigma\mathrm{K})} 
    \phi^{\mathrm{R}}_{m,(j\sigma'\mathrm{K})}.
    \label{Equation:expt}
\end{split}
\end{equation}
Using this equation, physical observables such as the particle number can be computed as shown in Eq.~\eqref{Equation:nt}, and other quantities, including $\langle c_{i\sigma} c_{j\sigma'} \rangle(t)$ and $\langle c_{i\sigma}^\dagger c_{j\sigma'}^\dagger \rangle(t)$, can be evaluated in a similar manner. 

As long as the Liouvillian spectrum has no eigenvalues with zero imaginary part, the steady state expectation value is independent of the initial state and is given by Eq.~\eqref{Equation:third}. Moreover, Eq.~\eqref{Equation:expt} allows us to compute the time evolution of observables for arbitrary initial states, provided that the overlaps $a_{nm}$ are known.

\section{Conversion from PBC to OBC \label{Appendix:conversion}}
We outline the procedure for changing the boundary condition from PBC to OBC along the $x$ direction.

First, we reorder the basis $\Psi_{\vb*{k}}$ into 
\begin{equation}
    \Psi_{\vb*{k}}' = \begin{pmatrix} \alpha_{\vb*{k}1 \mathrm{K}} &  \alpha_{\vb*{k}2 \mathrm{K}} & \alpha_{\vb*{k}1 \mathrm{B}}^{\dagger} &\alpha_{\vb*{k}2 \mathrm{B}}^{\dagger} \end{pmatrix}^T.
\end{equation}

Applying the unitary transformation 
\begin{equation}
    \tilde{U} = 
    \begin{pmatrix}
    U & 0 \\
    0 & U
    \end{pmatrix},
\end{equation} 
with $U$ defined by $\vb{C}_{\vb*{k}} = U\vb{C}'_{\vb*{k}}$, we rewrite the Liouvillian in the original electron basis as
\begin{eqnarray}
    &\mathcal{L} = \sum_{\vb*{k}}\vb{c}_{\vb*{k}}^\dagger \mathcal{L}^{\mathrm{c}}(\vb*{k})\vb{c}_{\vb*{k}},\\
    &\vb{c}_{\vb*{k}} =
     \begin{pmatrix} c_{\vb*{k}\uparrow \mathrm{K}} & c_{\vb*{k}\downarrow \mathrm{K}} & 
     c_{\vb*{k}\uparrow \mathrm{B}}^{\dagger} & 
     c_{\vb*{k}\downarrow \mathrm{B}}^{\dagger}
     \end{pmatrix}^T,
\end{eqnarray}
with $\vb{c}_{\vb*{k}}=\tilde{U}\Psi_{\vb*{k}}'$.

Next, we apply the Fourier transform along the 
$x$ direction
\begin{equation}
    \vb{c}_{\vb*{k}} = \frac{1}{\sqrt{L}}\sum_{j}e ^{-\ii k _{x}x_j}\vb{c}_{x_j,k _{y}},
    \label{Equation:fourier}
\end{equation}
where $L$ is the system size along the $x$ direction.
The real-space representation of the Liouvillian along the 
$x$ direction is obtained as follows:
\begin{eqnarray}
   & \mathcal{L}^{\mathrm{c}}(k_y)[x_i,x_j]=1/L \sum_{k _{x}}e ^{\ii k _{x} (x_i - x_j)}\mathcal{L}^{\mathrm{c}}(\vb*{k}),\\
   &\mathcal{L} = \sum_{k _{y}}\sum_{x_i,x_j}\vb{c}_{x_i,k_y}^\dagger\mathcal{L}^{\mathrm{c}}(k_y)[x_i,x_j] \vb{c}_{x_j,k_y}.
\end{eqnarray}

To obtain the mean-field Liouvillian under OBC, we eliminate matrix elements of $\mathcal{L}^{\mathrm{c}}(k_y)[x_i, x_j]$ that correspond to hopping across the boundaries. Since the hopping amplitude decays with distance, we define $l_{\mathrm{max}}$ as the maximum hopping range beyond which all hopping amplitudes fall below a given threshold $t_{\mathrm{lim}}$. All matrix elements corresponding to hopping across the boundaries within this range are discarded. In this way, we construct the Liouvillian under OBC.
In all the numerical calculations presented in this work, the threshold is
set to $t_{\mathrm{lim}} = 0.005$.

This procedure corresponds to discarding matrix elements of the Hermitian matrices $H$ and $M^{\mathrm{g(l)}}$ that connect sites across the boundaries. Since Hermiticity is preserved in the truncation, the form of Eq.~\eqref{Equation:Liouvillian_vec} remains unchanged, and the trace-preserving property of the Liouvillian is maintained. Therefore, the discussions in Appendix~\ref{Appendix:bogoliubov} and Appendix~\ref{Appendix:formula} can be applied to the open boundary case without modification.

\section{Correspondence between a quadratic Liouvillian and coupled Hatano-Nelson chains\label{Appendix:nhcse}}
\begin{figure}[htbp]
\includegraphics[width=0.35\textwidth]{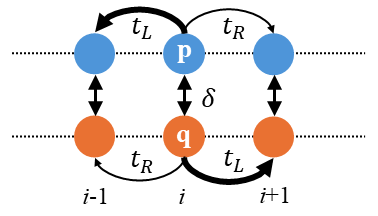}
\caption{
Coupled Hatano-Nelson chains. The blue and orange circles represent sites in sublattices p and q. We set hopping amplitude to be $t_{\mathrm{L}}> t_{\mathrm{R}}>0$. The parameter $\delta$ represents the coupling between two Hatano-Nelson chains. 
\label{Figure:coupledHN}
}
\end{figure}

We show correspondence between a quadratic Liouvillian and coupled Hatano-Nelson chains, a representative model exhibiting the non-Hermitian critical skin effect~\cite{Yokomizo_PRB2021,Li_NatComm2020}.

The Hatano-Nelson model~\cite{Hatano_PRB1997,Hatano_PRB1998,Hatano_PRL1996} describes a one-dimensional chain with asymmetric hopping. When two such chains with opposite hopping asymmetries are coupled as shown in Fig.~\ref{Figure:coupledHN}, the resulting system serves as a minimal model for the critical skin effect, capturing essential features such as system-size dependent spectra and scale-free localization. The model is described by the following Hamiltonian:
\begin{equation}
\begin{split}
    H &=   \sum_{i}(
    t_{\mathrm{L}}c_{i,\mathrm{p}}^\dagger c_{i+1,\mathrm{p}} +t_{\mathrm{R}}c_{i+1,\mathrm{p}}^\dagger c_{i,\mathrm{p}} \\
    &+t_{\mathrm{L}}c_{i+1,\mathrm{q}}^\dagger c_{i,\mathrm{q}}
    +t_{\mathrm{R}}c_{i,\mathrm{q}}^\dagger c_{i+1,\mathrm{q}}\\
    &+\delta c_{i,\mathrm{p}}^\dagger c_{i,\mathrm{q}} + \delta c_{i,\mathrm{q}}^\dagger c_{i,\mathrm{p}}
    ),
\end{split}
\end{equation}
where $c_{i,\mathrm{p}}^\dagger$ ($c_{i,\mathrm{q}}^\dagger$) and $c_{i,\mathrm{p}}$ ($c_{i,\mathrm{q}}$) are fermionic creation and annihilation operators on sublattice p (q) in the unit cell at site~$i$. We assume that $t_\mathrm{L}$ and $t_\mathrm{R}$ are real. 
Under PBC, the Hamiltonian can be expressed as
\begin{equation}
    \begin{split}
    &H=\sum_{k}
    \begin{pmatrix}
            c_{k,\mathrm{p}}^\dagger\\
            c_{k,\mathrm{q}}^\dagger
    \end{pmatrix}^T
    \begin{pmatrix}
        h_{\mathrm{HN}}(k) &\delta\\
        \delta & h^*_{\mathrm{HN}}(k)
    \end{pmatrix}
    \begin{pmatrix}
        c_{k,\mathrm{p}}\\
        c_{k,\mathrm{q}}
    \end{pmatrix}, \\
    & h_{\mathrm{HN}}(k)=t_{\mathrm{L}}e^{\ii k} + t_{\mathrm{R}}e^{-\ii k}.
    \end{split}    
\end{equation}
A key feature of the model is the coupling~$\delta$ between two modes with oppositely winding spectra.

On the other hand, a translationally invariant GKSL equation with single-particle gain and loss jump operators yields the following quadratic Liouvillian:
\begin{equation}
    \begin{split}
        &\mathcal{L}|_{\eta=1} = \sum_k \vb{c}_{k}^\dagger \vb{L}_k \vb{c}_{k} 
        +\mathrm{const} , \\
        &\vb{L}_k = 
        \begin{pmatrix}
            H_{k} + \ii (M^{\mathrm{g}}_{k})^T - \ii M^{\mathrm{l}}_k & 2\ii (M^{\mathrm{g}}_{k})^T \\
            2\ii M^{\mathrm{l}}_{k} & H_{k} - \ii (M^{\mathrm{g}}_{k})^T + \ii M^{\mathrm{l}}_{k}
        \end{pmatrix},\\
        &\vb{c}_k = \begin{pmatrix}
            c_{k\uparrow\mathrm{K}}&c_{k\downarrow\mathrm{K}}&c_{k\uparrow\mathrm{B}}^\dagger&c_{k\downarrow\mathrm{B}}^\dagger
        \end{pmatrix}.
    \end{split}
\end{equation}
Here, $H_k$ denotes the Bloch Hamiltonian of the system, and $M_k^{\mathrm{g}(\mathrm{l})} = 1/L \sum_{i,j} e^{-\ii(i-j)k} M_{i,j}^{\mathrm{g}(\mathrm{l})}$ are the gain and loss matrices in momentum space.
When the ket and bra sectors of $\vb{L}_k$ exhibit spectral winding, the Hermiticity of $H_k$ and $M_k$ ensures that the corresponding winding numbers have opposite signs.
Moreover, the ket and bra sectors are coupled via the matrices $(M_k^{\mathrm{g}})^T$ and $M_k^{\mathrm{l}}$.

Thus, the quadratic Liouvillian has a structural correspondence to the coupled Hatano–Nelson chains, which exhibit system-size dependence in the spectrum and localization length.
This correspondence suggests the system-size dependence of the Liouvillian skin effect.

\section{Transformation matrices acting on the damping matrix \texorpdfstring{$X$}{X}\label{Appendix:symmetry}}
Here, we examine how the damping matrix $X$ behaves under symmetry operations. We consider the transformation
\begin{equation}
    c_{m\vb*{k}\sigma} = M_{\sigma'\sigma}c_{\vb*{k}\sigma'},
\end{equation}
where $m\vb*{k}$ denotes the momentum label after the transformation.
Under this operation, the operators in the ket and bra spaces transform as follows:
\begin{equation}
\begin{split}
    &c_{m\vb*{k}\sigma\mathrm{K}} = M_{\sigma'\sigma}(c_{\vb*{k}\sigma'}\otimes \1)=  M_{\sigma'\sigma}c_{\vb*{k}\sigma'\mathrm{K}},\\
    &c_{m\vb*{k}\sigma\mathrm{B}} = M_{\sigma'\sigma}^*\eta(\1\otimes c_{\vb*{k}\sigma'}^*) = M_{\sigma'\sigma}^*c_{\vb*{k}\sigma'\mathrm{B}}.
\end{split} 
\end{equation}
Therefore, the Nambu basis 
$\vb{c}_{\vb*{k}} = \begin{pmatrix}
            c_{\vb*{k}\uparrow\mathrm{K}}&c_{\vb*{k}\downarrow\mathrm{K}}&c_{\vb*{k}\uparrow\mathrm{B}}^\dagger&c_{\vb*{k}\downarrow\mathrm{B}}^\dagger\end{pmatrix}$
transforms under the matrix
\begin{equation}
    \tilde{M} = \begin{pmatrix}
        M&\bm{0}\\\bm{0}&M
    \end{pmatrix}
\end{equation}
as
\begin{equation}
    \vb{c}_{m\vb*{k}} = \tilde{M}\vb{c}_{\vb*{k}}.
\end{equation}

As a result, the matrix $\vb{L}$ transforms as
\begin{equation}
    \vb{L} \mapsto \tilde{M}\vb{L}\tilde{M}^{-1},
\end{equation}
and the damping matrix $X$ transforms as
\begin{equation}
    X \mapsto MXM^{-1}.
\end{equation}

\section{
The Liouvillian under the mean-field approximation
\label{Appendix:meanfield}
}
We show the Liouvillian under the mean-field approximation possesses the following properties:
(i) the CPTP of the Liouvillian are preserved
under the approximation;
(ii) the steady state remains the same Gibbs distribution as that of the original Liouvillian due to the detailed balance condition; and
(iii) the particle number is effectively conserved, as confirmed by numerical simulations.

We begin by showing that (i) the CPTP of the Liouvillian is preserved under the mean-field approximation.
We consider the specific jump operator
\begin{equation}
    L_{12}^{\vb*{k}} =  \sqrt{\Gamma_{12}^{\vb*{k}}}\,\alpha_{\vb*{k}1}^\dagger \alpha_{\vb*{k}2},
\end{equation}
whose contribution to the dissipator in the GKSL equation is given by:
\begin{equation}
\begin{split}
    &L_{12}^{\vb*{k}}\otimes L_{12}^{\vb*{k}*} - \frac{1}{2}\left[L_{12}^{\vb*{k}\dagger} L_{12}^{\vb*{k}}\otimes \1 + \1 \otimes (L_{12}^{\vb*{k}\dagger} L_{12}^{\vb*{k}})^*\right] \\
    &= \Gamma_{12}^{\vb*{k}} \Big[
    \alpha_{\vb*{k}1 \mathrm{K}}^{\dagger} \alpha_{\vb*{k}2 \mathrm{K}} \alpha_{\vb*{k}1 \mathrm{B}}^{\dagger} \alpha_{\vb*{k}2 \mathrm{B}} \\
    &\quad - \frac{1}{2} \left(
    \alpha_{\vb*{k}2 \mathrm{K}}^{\dagger} \alpha_{\vb*{k}1 \mathrm{K}} \alpha_{\vb*{k}1 \mathrm{K}}^{\dagger} \alpha_{\vb*{k}2 \mathrm{K}}
    + \alpha_{\vb*{k}2 \mathrm{B}}^{\dagger} \alpha_{\vb*{k}1 \mathrm{B}} \alpha_{\vb*{k}1 \mathrm{B}}^{\dagger} \alpha_{\vb*{k}2 \mathrm{B}}
    \right)
    \Big].
\end{split}
\end{equation}

Applying the mean-field approximation, we obtain
\begin{equation}
\begin{split}
    &\Gamma_{12}^{\vb*{k}} f(-\varepsilon_{\vb*{k}1})\!\left[
    - \alpha_{\vb*{k}2 \mathrm{K}} \alpha_{\vb*{k}2 \mathrm{B}}
    - \frac{1}{2} \!\left(
    \alpha_{\vb*{k}2 \mathrm{K}}^\dagger \alpha_{\vb*{k}2 \mathrm{K}}
    + \alpha_{\vb*{k}2 \mathrm{B}}^\dagger \alpha_{\vb*{k}2 \mathrm{B}}
    \right)
    \right] \\
    &+ \Gamma_{12}^{\vb*{k}} f(\varepsilon_{\vb*{k}2})\! \left[
    \alpha_{\vb*{k}1 \mathrm{K}}^\dagger \alpha_{\vb*{k}1 \mathrm{B}}^\dagger
    - \frac{1}{2} \!\left(
    \alpha_{\vb*{k}1 \mathrm{K}} \alpha_{\vb*{k}1 \mathrm{K}}^\dagger
    + \alpha_{\vb*{k}1 \mathrm{B}} \alpha_{\vb*{k}1 \mathrm{B}}^\dagger
    \right)
    \right].
\end{split} \label{lind}
\end{equation}

Using Eq.~\eqref{Equation:KB}, the operators are expressed in tensor-product form:
\begin{equation}
\begin{split}
    &\Gamma_{12}^{\vb*{k}} f(-\varepsilon_{\vb*{k}1}) \Big[
    - (\alpha_{\vb*{k}2} \otimes \1)\, \eta\, (\1 \otimes \alpha_{\vb*{k}2})^* \\
    & - \frac{1}{2} \left(
    (\alpha_{\vb*{k}2}^\dagger \otimes \1) (\alpha_{\vb*{k}2} \otimes \1)
    + (\1 \otimes \alpha_{\vb*{k}2}^\dagger)^*\, \eta\, \eta\, (\1 \otimes \alpha_{\vb*{k}2})^*
    \right)
    \Big] \\
    &+ \Gamma_{12}^{\vb*{k}} f(\varepsilon_{\vb*{k}2}) \Big[
    (\alpha_{\vb*{k}1}^\dagger \otimes \1) (\1 \otimes \alpha_{\vb*{k}1}^\dagger)^* \eta \\
    &- \frac{1}{2} \left(
    (\alpha_{\vb*{k}1} \otimes \1) (\alpha_{\vb*{k}1}^\dagger \otimes \1)
    + \eta (\1 \otimes \alpha_{\vb*{k}1})^* (\1 \otimes \alpha_{\vb*{k}1}^\dagger)^* \eta
    \right)
    \Big].
\end{split}
\end{equation}

Projecting onto the $\eta = 1$ subspace (see Appendix~\ref{Appendix:bogoliubov}) and undoing the vectorization, we obtain
\begin{equation}
\begin{split}
    &\Gamma_{12}^{\vb*{k}} f(-\varepsilon_{\vb*{k}1}) \left(
    \alpha_{\vb*{k}2} \rho \alpha_{\vb*{k}2}^\dagger
    - \frac{1}{2} \left\{
    \alpha_{\vb*{k}2}^\dagger \alpha_{\vb*{k}2}, \rho
    \right\}
    \right) \\
    &+ \Gamma_{12}^{\vb*{k}} f(\varepsilon_{\vb*{k}2}) \left(
    \alpha_{\vb*{k}1}^\dagger \rho \alpha_{\vb*{k}1}
    - \frac{1}{2} \left\{
    \alpha_{\vb*{k}1} \alpha_{\vb*{k}1}^\dagger, \rho
    \right\}
    \right),
\end{split}
\end{equation}
which corresponds to the dissipative part of a GKSL equation with jump operators
\begin{equation}
   L_{12}^{\vb*{k}(1)} = \sqrt{\Gamma_{12}^{\vb*{k}} f(\varepsilon_{\vb*{k}2})} \, \alpha_{\vb*{k}1}^\dagger, \quad
   L_{12}^{\vb*{k}(2)} = \sqrt{\Gamma_{12}^{\vb*{k}} f(-\varepsilon_{\vb*{k}1})} \, \alpha_{\vb*{k}2}.
\end{equation}

Similarly, other jump operators are also decomposed into two jump operators under the mean-field approximation, as follows:

\begin{equation}
\begin{aligned}
   L_{21}^{\vb*{k}} &=  \sqrt{\Gamma_{21}^{\vb*{k}}}\alpha_{\vb*{k}2}^\dagger \alpha_{\vb*{k}1}
   \ \rightarrow \
   \begin{cases}
      L_{21}^{\vb*{k}(1)} = \sqrt{\Gamma_{21}^{\vb*{k}} f(\varepsilon_{\vb*{k}1})} \alpha_{\vb*{k}2}^\dagger, \\
      L_{21}^{\vb*{k}(2)} = \sqrt{\Gamma_{21}^{\vb*{k}} f(-\varepsilon_{\vb*{k}2})} \alpha_{\vb*{k}1},
   \end{cases} \\[1ex]
   L_{11}^{\vb*{k}} &= \sqrt{\dfrac{\gamma_{\mathrm{d}}}{2}}\, \alpha_{\vb*{k}1}^\dagger \alpha_{\vb*{k}1}
   \ \rightarrow \
   \begin{cases}
      L_{11}^{\vb*{k}(1)} = \sqrt{\dfrac{\gamma_{\mathrm{d}}}{2} f(\varepsilon_{\vb*{k}1})} \alpha_{\vb*{k}1}^\dagger, \\
      L_{11}^{\vb*{k}(2)} = \sqrt{\dfrac{\gamma_{\mathrm{d}}}{2} f(-\varepsilon_{\vb*{k}1})} \alpha_{\vb*{k}1},
   \end{cases} \\[1ex]
   L_{22}^{\vb*{k}} &= \sqrt{\dfrac{\gamma_{\mathrm{d}}}{2}}\alpha_{\vb*{k}2}^\dagger \alpha_{\vb*{k}2}
   \ \rightarrow \
   \begin{cases}
      L_{22}^{\vb*{k}(1)} = \sqrt{\dfrac{\gamma_{\mathrm{d}}}{2} f(\varepsilon_{\vb*{k}2})} \alpha_{\vb*{k}2}^\dagger, \\
      L_{22}^{\vb*{k}(2)} = \sqrt{\dfrac{\gamma_{\mathrm{d}}}{2} f(-\varepsilon_{\vb*{k}2})} \alpha_{\vb*{k}2}.
   \end{cases}
\end{aligned}
\end{equation}
This decomposition ensures that the Liouvillian retains the GKSL form under the mean-field approximation, which guarantees its CPTP.

Next, we show that (ii) the steady state remains the same Gibbs distribution as that of the original Liouvillian due to the detailed balance condition.
Let $\Gamma_{\vb*{k}1}^{\mathrm{g}}$ and $\Gamma_{\vb*{k}1}^{\mathrm{l}}$ denote the effective gain and loss rates for the quasiparticle $\alpha_{\vb*{k}1}$, respectively. These rates are given by
\begin{equation}
\begin{split}
    \Gamma_{\vb*{k}1}^{\mathrm{g}} &= \Gamma_{12}^{\vb*{k}} f(\varepsilon_{\vb*{k}2}) + \dfrac{\gamma_{\mathrm{d}}}{2} f(\varepsilon_{\vb*{k}1}), \\
    \Gamma_{\vb*{k}1}^{\mathrm{l}} &= \Gamma_{21}^{\vb*{k}} f(-\varepsilon_{\vb*{k}2}) + \dfrac{\gamma_{\mathrm{d}}}{2} f(-\varepsilon_{\vb*{k}1}).
\end{split}
\end{equation}
These satisfy the detailed balance condition
\begin{equation}
    \dfrac{\Gamma_{\vb*{k}1}^{\mathrm{g}}}{\Gamma_{\vb*{k}1}^{\mathrm{l}}} = e^{-\beta \varepsilon_{\vb*{k}1}}.
\end{equation}

Similarly, for the quasiparticle $\alpha_{\vb*{k}2}$, the gain and loss rates are given by
\begin{equation}
\begin{split}
    \Gamma_{\vb*{k}2}^{\mathrm{g}} &= \Gamma_{21}^{\vb*{k}} f(\varepsilon_{\vb*{k}1}) + \dfrac{\gamma_{\mathrm{d}}}{2} f(\varepsilon_{\vb*{k}2}), \\
    \Gamma_{\vb*{k}2}^{\mathrm{l}} &= \Gamma_{12}^{\vb*{k}} f(-\varepsilon_{\vb*{k}1}) + \dfrac{\gamma_{\mathrm{d}}}{2} f(-\varepsilon_{\vb*{k}2}),
\end{split}
\end{equation}
and they also satisfy the detailed balance condition
\begin{equation}
    \dfrac{\Gamma_{\vb*{k}2}^{\mathrm{g}}}{\Gamma_{\vb*{k}2}^{\mathrm{l}}} = e^{-\beta \varepsilon_{\vb*{k}2}}.
\end{equation}
Since the detailed balance condition is satisfied, the steady state of the Liouvillian after the mean-field approximation remains the same as before the approximation, namely, the Gibbs equilibrium state of the Hamiltonian $H$.

\begin{figure}[htbp]
\includegraphics[width=0.45\textwidth]{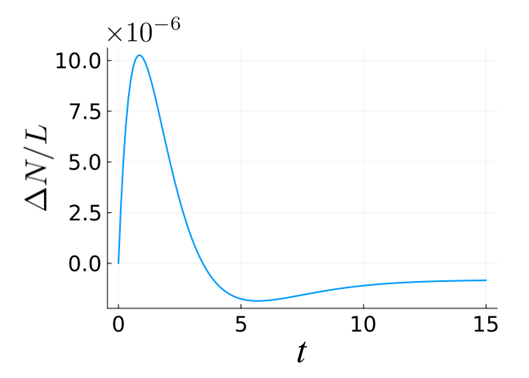}
\caption{
Time evolution of the deviation in the total particle number from its initial value, normalized by the system size $L$.
 To ensure that the bulk particle density matches between the initial and steady states, the initial chemical potential was slightly adjusted to $\mu = 0.9998945$, while the time evolution was computed using $\mu = 1$.
All data are obtained for $t_{\mathrm{h}} = 1$, $\gamma = 1$, $\gamma_{\mathrm{d}} = 1$, $k_y = 3\pi/4$, $T=0.5$, and $L = 50$.
\label{Figure:preserveN}
}
\end{figure}

Finally, we numerically demonstrate that (iii) the particle number is effectively conserved.  We compute the time evolution of the deviation in the total particle number from its initial value, as shown in Fig.~\ref{Figure:preserveN}. To enable direct comparison with Fig.~\ref{Figure:accumulation}, we use identical parameters and normalize the deviation by the system size $L$. At each time step, the normalized change $\Delta N / L$ is smaller than the accumulated particle number near the boundary in Fig.~\ref{Figure:accumulation} by a factor of $10^{-1} \sim 10^{-2}$, confirming that the particle number is effectively conserved under the mean-field approximation.

\section{
Symmetry Constraints with general antisymmetric SOC\label{Appendix:otherSOC}
}
We present symmetry constraints for generic antisymmetric SOC and identify the direction of the applied magnetic field that breaks these constraints and induces the skin effect.
To this end, we classify the possible symmetries 
of the Hamiltonian for each element of the group 
\( D_{2h} \oplus \mathcal{T}D_{2h} \), generated by the mirror symmetries
\( M_x, M_y, M_z \) and the time-reversal symmetry \( \mathcal{T} \),
all of which flip the momentum $k_x$.

First, we divide the Bloch Hamiltonian into the following three contributions:
\begin{equation}
    H(\boldsymbol{k}) = h_{\rm hop}(\boldsymbol{k}) + h_{\rm SOC}(\boldsymbol{k}) + h_{\rm Zeeman}(\boldsymbol{k}),
\end{equation}
where \( h_{\rm hop} \), \( h_{\rm SOC} \), and \( h_{\rm Zeeman} \) represent the hopping, SOC, Zeeman terms, respectively. 
For each of these components, we consider how they transform under the symmetry operations.

For any element \( s \) of the group \( D_{2h} \oplus \mathcal{T}D_{2h} \),
the transformation of each component of the Hamiltonian can be expressed as
\begin{equation}
    s \, h_{c}(\boldsymbol{k}) \, s^{-1}
    = {\rm sgn}_{c}(s)\, h_{c}(s\boldsymbol{k}),
    \qquad c = {\rm hop, SOC, Zeeman},
\end{equation}
where \( s\boldsymbol{k} \) denotes the momentum transformed under the operation \( s \),
and \( {\rm sgn}_{c}(s) = \pm 1 \) represents whether the corresponding term 
\( h_{c} \) is even (\(+1\)) or odd (\(-1\)) under the symmetry operation \( s \).

In our setup, the hopping term always satisfies
\begin{equation}
    {\rm sgn}_{\rm hop}(s) = +1.
\end{equation}
Therefore, for the entire Hamiltonian to preserve the symmetry $s$, 
the SOC and Zeeman terms must satisfy
\begin{equation}
    {\rm sgn}_{\rm SOC}(s) = {\rm sgn}_{\rm Zeeman}(s) = +1.
\end{equation}

Next, for each element \( s \) of the group \( D_{2h} \oplus \mathcal{T}D_{2h} \), 
we list the corresponding values of \({\rm sgn}_{\rm SOC}(s)\) in Table~\ref{table:sgnSOC}.
Since we are considering an antisymmetric SOC, we have 
\({\rm sgn}_{\rm SOC}(\mathcal{I}) = -1\).
Moreover, because time-reversal operation
reverses both momentum and spin, 
it follows that \({\rm sgn}_{\rm SOC}(\mathcal{T}) = +1\).
In addition, from the relation
\begin{equation}
    M_z = \mathcal{I} M_x M_y,
\end{equation}
once the values of \({\rm sgn}_{\rm SOC}(M_x)\) and \({\rm sgn}_{\rm SOC}(M_y)\) are fixed,
all the remaining \({\rm sgn}_{\rm SOC}(s)\) values are determined accordingly.
Therefore, the possible sets of \({\rm sgn}_{\rm SOC}(s)\) can be classified into four categories as shown in Table~\ref{table:sgnSOC}.

Representative examples of antisymmetric SOCs in two-dimensional materials --- Rashba- and Dresselhaus-types --- can be classified as follows:

\begin{align}
    &h_{\rm SOC}^{\rm (Rashba)} \propto (k_x \sigma_y - k_y \sigma_x) 
        \rightarrow{\text{(b)}},\\
    &h_{\rm SOC}^{\rm (Dresselhaus)} \propto (k_x \sigma_x - k_y \sigma_y) 
        \rightarrow{\text{(a)}}.
\end{align}

By identifying preserved symmetry (marked as $+$ in Table~\ref{table:sgnSOC}) for each antisymmetric SOC, one can determine the directions in which the Liouvillian skin effects are forbidden i.e., the directions in which the corresponding winding numbers are zero due to symmetry operations that flip the wave vector

For instance, in the case of the Rashba-type SOC [see row (b) in Table~\ref{table:sgnSOC}], 
the system is invariant under the operations of $M_x$ and $T M_y$ both of which flip  $k_x$.
As discussed in Sec.~\ref{sec: symm const}, 
these symmetry constraints ($M_x$ and $TM_y$) lead to vanishing winding number [Eq.~\eqref{eq:winding}] and thereby prohibit the Liouvillian skin effect.
The same argument can be applied to the other directions and to the other anti symmetric SOC as well.

Finally, we discuss the magnetic field that breaks the symmetry constraints prohibiting the Liouvillian skin effect.
For each element $s$ of the group $D_{2h} \oplus \mathcal{T}D_{2h}$, 
the corresponding values of ${\rm sgn}_{\rm Zeeman}(s)$ are listed in Table~\ref{table:sgnZeeman}. 
In Table~\ref{table:sgnZeeman}, a negative sign $(-)$ indicates that the corresponding symmetry is broken by the applied Zeeman field. 
Therefore, by applying a magnetic field along an appropriate direction, 
one can break the symmetry constraints on the winding number, thereby inducing the Liouvillian skin effect. 
For the Rashba-type SOC, as an example, applying a Zeeman field $H_y$ breaks the $M_x$ and $TM_y$ symmetries that prohibit a non zero winding number along the $x$ direction [see row (b) of Table~II and Table~III]. Thereby, applying the magnetic field in the $y$ direction to systems with the Rashba-type SOC leads to Liouvillian skin effect (see Fig.~\ref{Figure:localization}).

When a single magnetic-field component cannot break all the symmetry constraints, 
one can apply magnetic fields along multiple directions to break them.
For instance, in the presence of the SOC shown in Table~\ref{table:sgnSOC}(c), 
the symmetries that prohibit winding along the $k_x$ direction are $\mathcal{T}M_y$ and $\mathcal{T}\mathcal{I}M_x$. 
From Table~\ref{table:sgnZeeman}, we find that both of these symmetries can be broken only when the Zeeman fields $H_x$ and $H_y$ are applied simultaneously. 

In addition to in-plane fields, a Zeeman field along the $z$ direction can also 
break symmetry constraints.
For example, in the presence of the chiral-type SOC, symmetry constraints of $M_x$ and $\mathcal{I}M_y$ prohibit a non zero winding number along the $x$ direction [see row (d) in Table II]. As shown in Table III, applying a magnetic field $H_z$ breaks these symmetries, indicating the emergence of the Liouvillian skin effect.

\begin{table*}[]
\centering
\caption{
Signs of ${\rm sgn}_{\rm SOC}(s)$ for each element of the group 
\( D_{2h} \oplus \mathcal{T}D_{2h} \).
For each symmetry operation $s$, the components of momentum 
$\boldsymbol{k}=(k_x,k_y)$ that are inverted are listed in the second row. 
Here, $M_z$, $\mathcal{I}M_z$, $\mathcal{T}M_z$, $\mathcal{I}\mathcal{T}M_z$, and $\mathcal{I}\mathcal{T}$ 
are not shown since they do not invert either $k_x$ or $k_y$.
The four rows below correspond to different types of antisymmetric SOC characterized by 
$[{\rm sgn}_{\rm SOC}(M_x),{\rm sgn}_{\rm SOC}(M_y)] 
= [-1,-1], [+1,+1], [-1,+1],$ and $[+1,-1]$, respectively. 
}
\label{table:sgnSOC}
\begin{ruledtabular}
\begin{tabular}{@{\hskip 1pt}c|cccccccccc@{\hskip 0.5pt}}
 & $\mathcal{I}$ & $\mathcal{T}$ & $M_x$ & $M_y$ &
$\mathcal{I}M_x$ & $\mathcal{I}M_y$ &
$\mathcal{T}M_x$ & $\mathcal{T}M_y$ &
$\mathcal{T}\mathcal{I}M_x$ & $\mathcal{T}\mathcal{I}M_y$ \\
\hline
inverted $\boldsymbol{k}$ & 
$k_x,k_y$ & $k_x,k_y$ & $k_x$ & $k_y$ &
$k_y$ & $k_x$ &
$k_y$ & $k_x$ &
$k_x$ & $k_y$ \\
\hline
(a)& $-$ & $+$ & $-$ & $-$ & $+$ & $+$ & $-$ & $-$ & $+$ & $+$ \\
(b)& $-$ & $+$ & $+$ & $+$ & $-$ & $-$ & $+$ & $+$ & $-$ & $-$ \\
(c)& $-$ & $+$ & $-$ & $+$ & $+$ & $-$ & $-$ & $+$ & $+$ & $-$ \\
(d)& $-$ & $+$ & $+$ & $-$ & $-$ & $+$ & $+$ & $-$ & $-$ & $+$ \\
\end{tabular}
\end{ruledtabular}
\end{table*}

\begin{table*}[t]
\centering
\caption{Signs of ${\rm sgn}_{\rm Zeeman}(s)$ for each element of the group 
\( D_{2h} \oplus \mathcal{T}D_{2h} \).
For each symmetry operation $s$, the components of momentum 
$\boldsymbol{k}=(k_x,k_y)$ that are inverted are listed in the second row of the table. 
Here, $M_z$, $\mathcal{I}M_z$, $\mathcal{T}M_z$, $\mathcal{I}\mathcal{T}M_z$, and $\mathcal{I}\mathcal{T}$ 
are not shown since they do not invert either $k_x$ or $k_y$.
The three rows below correspond to the directions of the applied Zeeman field. 
}
\label{table:sgnZeeman}
\begin{ruledtabular}
\begin{tabular}{@{\hskip 1pt}c|cccccccccc@{\hskip 0.5pt}}
 & $\mathcal{I}$ & $\mathcal{T}$ & $M_x$ & $M_y$ &
 $\mathcal{I}M_x$ & $\mathcal{I}M_y$ &
 $\mathcal{T}M_x$ & $\mathcal{T}M_y$ &
 $\mathcal{T}\mathcal{I}M_x$ & $\mathcal{T}\mathcal{I}M_y$ \\
\hline
inverted $\boldsymbol{k}$ & 
$k_x,k_y$ & $k_x,k_y$ & $k_x$ & $k_y$ &
$k_y$ & $k_x$ &
$k_y$ & $k_x$ &
$k_x$ & $k_y$ \\
\hline
$H_x$ & $+$ & $-$ & $+$ & $-$ & $+$ & $-$ & $-$ & $+$ & $-$ & $+$ \\
$H_y$ & $+$ & $-$ & $-$ & $+$ & $-$ & $+$ & $+$ & $-$ & $+$ & $-$ \\
$H_z$ & $+$ & $-$ & $-$ & $-$ & $-$ & $-$ & $+$ & $+$ & $+$ & $+$ \\
\end{tabular}
\end{ruledtabular}
\end{table*}

\section{Breaking of the $\mathbb{Z}_2$ skin modes\label{Appendix:BreakingZ2}}

\begin{figure}[htbp]
\includegraphics[width=0.48\textwidth]{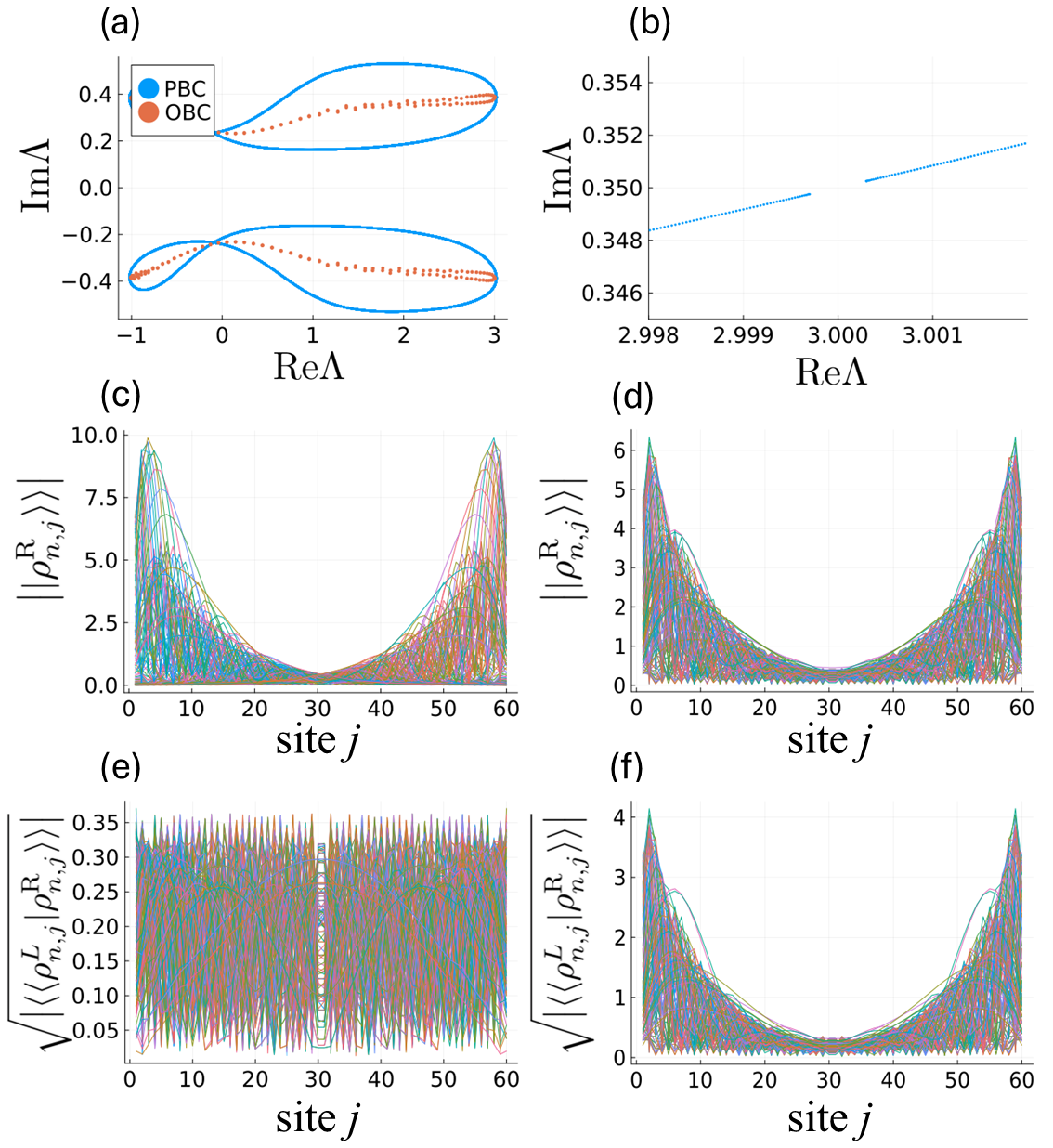}
\caption{
$\mathbb{Z}_2$ skin modes, and eigenvalues and eigenstates slightly away from $k_y=\pi$.
(a): Eigenvalues
of the Liouvillian under PBC in $y$ direction and OBC or PBC in $x$ direction. Panel (a) is
obtained for $k_y=\pi + 0.01$. (b): Magnified version of the panel (a).
(c) [(d)]: The weight of the right eigenstates obtained for $k_y = \pi$ [$k_y=\pi+0.001$]. 
Here, the weight is defined as
$\abs{|\rho_{n,j}^{\mathrm{R}}\rangle\rangle }
\equiv 
\sum_{\sigma=\uparrow,\downarrow;\;\tau=\mathrm{K,B}}
\bigl\lvert|\rho^{\mathrm{R}}_{n,j\sigma\tau}\rangle\!\rangle \bigr\rvert,
$ 
and $n$ is a label of the corresponding eigenvalue of the Liouvillian.
(e) [(f)]: The weight computed from both right and left eigenstates, using the same parameters as in panel (c) [(d)]. 
Here, the weight is defined as
$\sqrt{\abs{\langle\langle\rho_{n,j}^{L}|\rho_{n,j}^{\mathrm{R}}\rangle\rangle}  }
\equiv 
\sum_{\sigma=\uparrow,\downarrow;\;\tau=\mathrm{K,B}}
\sqrt{\abs{\langle\!\langle \rho^{\mathrm{L}}_{n,j\sigma\tau}\mid \rho^{\mathrm{R}}_{n,j\sigma\tau}\rangle\!\rangle\,}
}.
$
The data were obtained for $t_{\mathrm{h}}=1$, $\alpha=0.3$, $\mu=1$, $H_x=0$, $H_y=0$, $\gamma=1$, $\gamma_{\mathrm{d}}=0.2$, and $T=0.3$. The system size is $L=60$ for OBC and $L=5\times10^{4}$ for PBC.
\label{Figure:BreakingZ2}
}
\end{figure}

When the system has no magnetic field but possesses antisymmetric SOC, skin modes protected by time-reversal symmetry emerge at the boundaries for $k_y=0$ or $\pi$.
However, away from $k_y=0,\pi$
time-reversal symmetry is no longer preserved and the point gap topology becomes trivial.
To confirm this, we plot the Liouvillian spectrum for $k_y = \pi + 0.01$ in Figs.~\ref{Figure:BreakingZ2}(a) and \ref{Figure:BreakingZ2}(b).
As seen in Fig.~\ref{Figure:BreakingZ2}(b), 
the spectrum becomes disconnected and can be continuously deformed into a single point, indicating that the topology is trivialized.
In this situation, our numerical analysis indicates that the eigenstates remain localized despite the trivial topology.
These localized states are not skin modes since the localization persists even in the weight computed from the right and left eigenstates.
Specifically, 
Figs.~\ref{Figure:BreakingZ2}(c) and \ref{Figure:BreakingZ2}(e) display 
the weights of the right eigenstates and those of the products of the right and left eigenstates, respectively.
The weight of right eigenstates indicates the localization around the edges [Fig.~\ref{Figure:BreakingZ2}(c)], whereas such localization is absent in the weight computed from both the right and left eigenstates [Fig.~\ref{Figure:BreakingZ2}(e)].
This behavior reflects the defining feature of skin modes that the right and left eigenstates are localized at opposite boundaries.
In contrast, for $k_y = \pi + 0.01$, the localization observed in Fig.~\ref{Figure:BreakingZ2}(d) persists even in the weight computed from both the right and left eigenstates [see Fig.~\ref{Figure:BreakingZ2}(f)], indicating that the localized states are not skin modes.

Clarifying the origin of such boundary-localized modes in the topologically trivial phase is left for future work.

\section{Localization under OBC for both directions}
In the main text, we analyzed the Liouvillian skin effect by considering a one-dimensional subspace at fixed $k_y$ under PBC in the $y$ direction (see Sec.~\ref{sec: localization}). Here, we analyze the localized modes under OBC in both directions (full OBC), which are found to be consistent with the localization behavior discussed in Sec.~\ref{sec: localization}.

We start with the case where a magnetic field is applied along the $y$ direction.
As discussed in Sec.~\ref{sec: localization}, under PBC in both directions, the winding number for each fixed $k_y$ is finite
because the transverse field breaks the symmetry constraints in Eqs.~\eqref{eq: symm_Mx} and \eqref{eq: symm_TMy}.
Consequently, when OBC are imposed along the $x$ direction, skin modes appear.
In contrast, for fixed $k_x$, the following symmetry constraints enforce a vanishing winding number (see Tables \ref{table:sgnSOC} and \ref{table:sgnZeeman}.
\begin{equation}
\begin{aligned}
&M_{y}\, X(k_x,k_y)\, M_{y}^{-1} \;=\; X(k_x,-k_y),\\
&\mathcal{T}M_x \, X(k_x,k_y)^{T} (\mathcal{T}M_x)^{-1} \;=\; X(k_x,-k_y) .
\end{aligned}
\label{eq:ky_symm_constraints}
\end{equation}
Correspondingly, no localization occurs under OBC along the $y$ direction.
(If the magnetic field is instead applied along the $x$ direction, the situation is reversed: localization appears along the $y$ direction but not along the $x$ direction.)

\begin{figure}[htbp]
\includegraphics[width=0.48\textwidth]{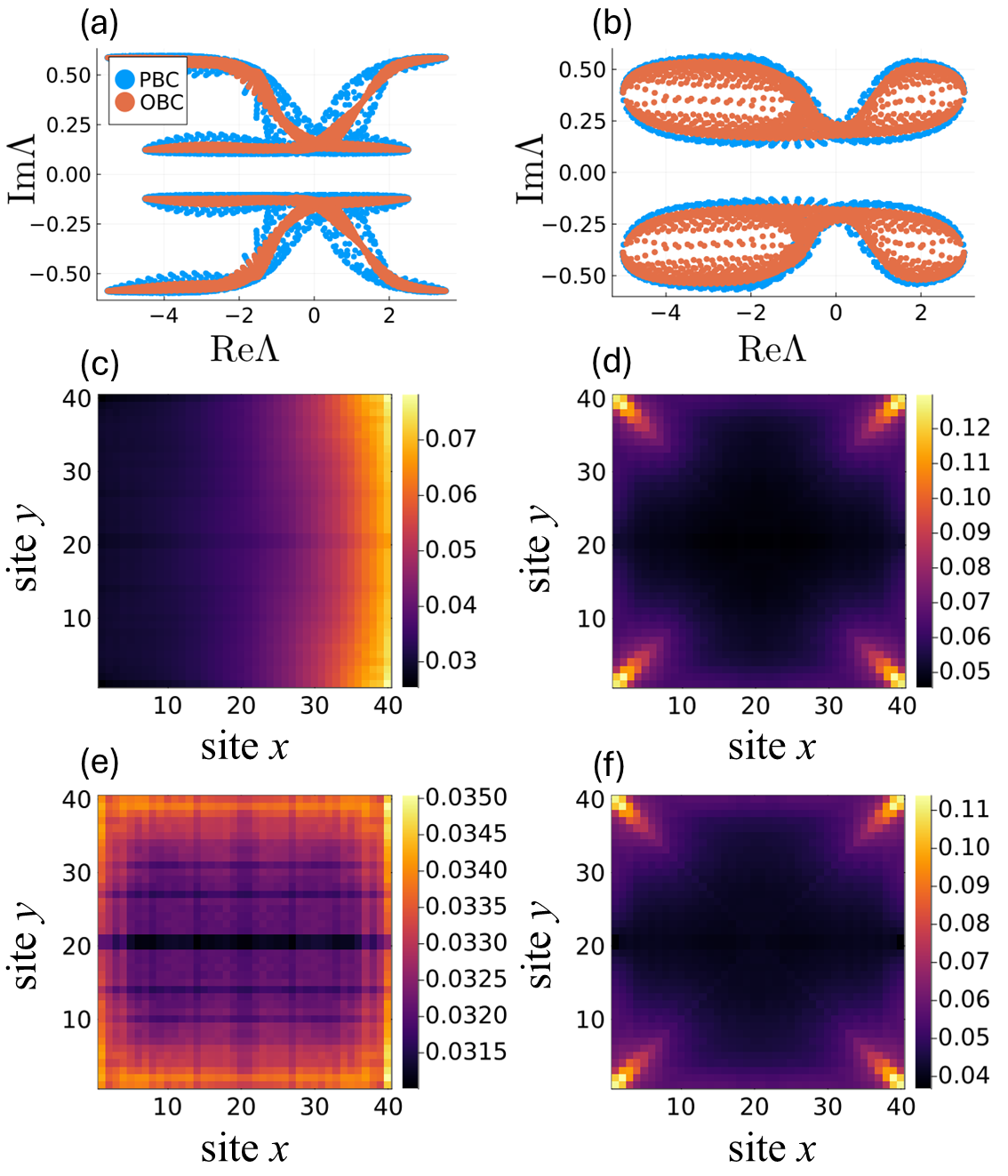}
\caption{
Spectrum and eigenstates under full OBC.
(a) and (b): Eigenvalues
of the Liouvillian.
(c) and (d): The weight of the right eigenstates  
$w_{j,\mathrm{R}}
\equiv 
\sum_{n;\;\sigma=\uparrow,\downarrow;\;\tau=\mathrm{K,B}}
\bigl\lvert|\rho^{\mathrm{R}}_{n,j\sigma\tau}\rangle\!\rangle \bigr\rvert/(L_xL_y).
$
(e) and (f): The weight computed from both the right and left eigenstates
$w_{j,\rm{LR}}
\equiv 
\sum_{n;\;\sigma=\uparrow,\downarrow;\;\tau=\mathrm{K,B}}
\sqrt{\langle\!\langle \rho^{\mathrm{L}}_{n,j\sigma\tau}\mid \rho^{\mathrm{R}}_{n,j\sigma\tau}\rangle\!\rangle\,
}/(L_xL_y)
$.
Panels (a), (c), and (e) [(b), (d), and (f)] are obtained for $H_y=1.0$ [$H_y=0$].
The data are obtained for $t_{\mathrm{h}}=1$, $\alpha=0.3$, $\mu=1$, $H_x = 0$,
$\gamma=1$, $\gamma_{\mathrm{d}}=0.2$, $T=0.3$, and $L_x=L_y=40$.
\label{Figure:FullOBC}
}
\end{figure}

For full OBC, based on the above behavior, we expect the skin effect to persist only along the $x$ direction, which is confirmed by numerical analysis under full OBC displayed in Fig.~\ref{Figure:FullOBC}.

Figure~\ref{Figure:FullOBC}(a) shows a clear spectral mismatch between PBC and OBC, indicating the emergence of the skin effect. 
Figure~\ref{Figure:FullOBC}(c) further presents the weight of the right eigenstate, which is localized along the $x$ direction while remaining delocalized along the $y$ direction.
In Fig.~\ref{Figure:FullOBC}(e)
it is found to be delocalized.
This delocalization arises from the left and right eigenstates localized at opposite boundaries, which supports the emergence of skin modes in Fig.~\ref{Figure:FullOBC}(c).

For the case without a magnetic field, at $k_y$ slightly deviated from $0$ or $\pi$, the system hosts localized modes. These localized modes are also observed in the weight of the left and right eigenstates, in contrast to the case of $k_y=0$ or $\pi$.
When open boundary conditions are imposed in both directions, we numerically confirm that skin modes with the same characteristic behavior become localized.
For full OBC, the weights of the right eigenstates and of the product of the right and left eigenstates exhibit similar localization [see Figs.~\ref{Figure:FullOBC}(d) and \ref{Figure:FullOBC}(f)]. These localized modes are considered not to be skin modes since the skin effect leads to the right and left eigenstates localized at opposite edge.

\clearpage

\bibliography{LSEinElectron}

\begin{thebibliography}{102}
\bibitem [{rel()}]{relaxNote}%
  \BibitemOpen
  \href@noop {} {}\bibinfo {note} {In a realistic two-dimensional electron system on a substrate, energy relaxation processes accompanied by momentum relaxation are generally present. In this work, however, we focus on interband energy transitions without momentum transfer for analytical convenience. Since the Liouvillian skin effect is a topological phenomenon, the inclusion of weak momentum relaxation processes merely acts as a perturbation to the topology, and the main results of this paper are expected to remain essentially unchanged.}\BibitemShut {Stop}%
\bibitem [{mom()}]{momentumNote}%
  \BibitemOpen
  \href@noop {} {}\bibinfo {note} {We introduce Fourier transforms in the ket and bra spaces separately: $c_{k\sigma} \otimes \1 = L^{-1/2} \sum_i e^{-\ii k x_i} \, c_{x_i\sigma} \otimes \1$ and $\1 \otimes c_{k\sigma}^{\ast} = L^{-1/2} \sum_i e^{+\ii k x_i} \; \1 \otimes c_{x_i\sigma}^{\ast}$, so we define $c_{k\sigma\mathrm{K}} \equiv c_{k\sigma} \otimes \1$ and $c_{k\sigma\mathrm{B}} \equiv \eta\bigl( \1 \otimes c_{k\sigma}^{\ast} \bigr)$. Here, $L$ denotes the system size.}\BibitemShut {Stop}%
\bibitem [{Nam()}]{Nambunote}%
  \BibitemOpen
  \href@noop {} {}\bibinfo {note} {In the standard notation used for superconductors, the momentum index of a hole typically has the opposite sign to that of a particle. However, in the present case, we have defined the Fourier transforms in the ket and bra spaces with opposite signs in the exponential factors. As a result, the particle and hole components share the same momentum index in our formulation.}\BibitemShut {Stop}%
\bibitem [{loc()}]{localizationNote}%
  \BibitemOpen
  \href@noop {} {}\bibinfo {note} {The correspondence between the winding number and the direction of localization depends on the convention of the Fourier transform. In this paper, we adopt the definition given in Eq.~\eqref{Equation:fourier}.}\BibitemShut {Stop}%
\bibitem [{win()}]{windingNote}%
  \BibitemOpen
  \href@noop {} {}\bibinfo {note} {Owing to the mirror reflection symmetry, the winding number vanishes, which can be confirmed as follows: $w(\Lambda_0) =\frac{1}{2\pi \ii}\int_{-\pi}^{\pi} dk_x\partial_{k_x}\ln \det [M X(-k_x,k_y)M^{-1} - \Lambda_0\1] =\frac{1}{2\pi \ii}\int_{-\pi}^{\pi} dk_x\partial_{k_x}\ln \det [X(-k_x,k_y) - \Lambda_0\1] =-w$. When the symmetry involves the transposition, the same argument holds since $\det[X^{\mathrm{T}}-\Lambda_0\1] = \det[X-\Lambda_0\1]$.}\BibitemShut {Stop}%
\bibitem [{Hy0()}]{Hy0Note}%
  \BibitemOpen
  \href@noop {} {}\bibinfo {note} {When $H_y$ is exactly zero, time-reversal symmetry enforces the degeneracy of the eigenvalues, which makes it difficult to extract the proper skin modes. In this work, we avoid this difficulty by introducing an extremely weak transverse magnetic field, typically $H_y = 10^{-6}$, which slightly lifts the degeneracy while leaving the structure of the spectrum essentially unchanged. When $k_y$ is slightly shifted, the spectrum becomes discontinuous and the system enters a topologically nontrivial phase. These effects are discussed in detail in Appendix~\ref{Appendix:BreakingZ2}.}\BibitemShut {Stop}%
\bibitem [{Par()}]{ParametersNote}%
  \BibitemOpen
  \href@noop {} {}\bibinfo {note} {To distinguish the temperature dependence governed by $\alpha$ and $H_y$ from that set by the energy scale of $t$ and $\mu$, we deliberately choose $\alpha$ and $H_y$ to be smaller than these parameters.}\BibitemShut {Stop}%
\bibitem [{Z2N()}]{Z2Note}%
  \BibitemOpen
  \href@noop {} {}\bibinfo {note} {A similar system-size dependence is also observed in the $\mathbb{Z}_2$ Liouvillian skin effects.}\BibitemShut {Stop}%
\bibitem [{pos()}]{positivesemiNote}%
  \BibitemOpen
  \href@noop {} {}\bibinfo {note} {The positive semidefiniteness of the matrix $M$ can be established as follows: for any vector $\vb*{x}$, $\vb*{x}^\dagger M^{\mathrm{g}(\mathrm{l})} \vb*{x} = \sum_{\mu} \frac{1}{2}\left| \sum_{i\sigma} D^{\mathrm{g}(\mathrm{l})}_{\mu,i\sigma} x_{i\sigma} \right|^2 \ge 0$.}\BibitemShut {Stop}%
\end{thebibliography}

\end{document}